\documentclass[acmlarge]{acmart}
\acmSubmissionID{9307}

\usepackage{hyperref}
\usepackage{hyperxmp}

\usepackage{threeparttable}
\usepackage{tablefootnote}

\def\sysname{\textsc{USpeech}\xspace}

\usepackage{tcolorbox}
\usepackage{colortbl}
\usepackage{fp}
\usepackage{romannum}
\usepackage[ruled]{algorithm2e}
\usepackage{algpseudocode}
\usepackage{enumerate}
\usepackage{enumitem}
\usepackage[english]{babel}
\usepackage{blindtext}
\usepackage[caption=false,font=footnotesize,subrefformat=parens]{subfig}
\usepackage{amsmath}

\usepackage{amssymb}
\usepackage{xspace}
\usepackage{multirow}
\usepackage{threeparttable}
\usepackage{float}
\usepackage{flafter}
\usepackage{comment}
\usepackage[skip=12pt]{caption}
\usepackage{graphicx}
\usepackage{soul}
\usepackage{color}
\usepackage{subfig}
\usepackage{bbding}
\usepackage{pifont}

\ifodd 0
\newcommand{\rev}[1]{#1}
\newcommand{\mrev}[1]{{\color{blue}#1}}
\newcommand{\com}[1]{\textbf{\color{red}(COMMENT: #1)}}
\newcommand{\todo}[1]{\textbf{{\color{orange}(TODO: #1)}}}
\else
\newcommand{\rev}[1]{#1}
\newcommand{\mrev}[1]{#1}
\newcommand{\com}[1]{}
\newcommand{\todo}[1]{}
\fi

\usepackage{tikz}
\newcommand*\emptycirc[1][0.618ex]{\tikz\draw (0,0) circle (#1);}

\newcommand*\halfcirc[1][0.618ex]{%
  \begin{tikzpicture}
  \draw[fill] (0,0)-- (90:#1) arc (90:270:#1) -- cycle ;
  \draw (0,0) circle (#1);
  \end{tikzpicture}}
  
\newcommand*\quartercirc[1][0.618ex]{%
\begin{tikzpicture}
\draw[fill] (0,0) -- (90:#1) arc (90:180:#1) -- cycle ;
\draw (0,0) circle (#1);
\end{tikzpicture}}

\newcommand*\threequartercirc[1][0.618ex]{%
  \begin{tikzpicture}
  \draw[fill] (0,0) -- (90:#1) arc (90:360:#1) -- cycle ;
  \draw (0,0) circle (#1);
  \end{tikzpicture}}
\newcommand*\fullcirc[1][0.618ex]{\tikz\fill (0,0) circle (#1);} 

\def\fig{Fig.\xspace}

\def\tab{Tab.\xspace}

\def\ie{{\textit{i.e.}\xspace}} 
\def\eg{{\textit{e.g.}\xspace}}

\def\etc{{\textit{etc.}\xspace}}
\def\vs{{\textit{v.s.}\xspace\xspace}}

\newcommand{\head}[1]{{\noindent \textbf{#1:}}}

\graphicspath{{./figures/}}
\graphicspath{{./pdf/}}
\DeclareGraphicsExtensions{.pdf,.jpeg,.png}

\usepackage{fp}

\newlength\maxlentime
\newcommand\pesqtimebar[3][red!20]{%
  \FPeval\result{round((#3-0)/#2:4)}%
  \rlap{\textcolor{#1}{\hspace*{\dimexpr-\tabcolsep+.5\arrayrulewidth}%
        \rule[-.05\ht\strutbox]{\result\maxlentime}{.95\ht\strutbox}}}%
  \makebox[\dimexpr\maxlentime-0.2\tabcolsep+\arrayrulewidth][c]{#3}}

\newcommand\stoitimebar[3][NavyBlue!20]{%
\FPeval\result{round((#3-0)/#2:4)}%
\rlap{\textcolor{#1}{\hspace*{\dimexpr-\tabcolsep+.5\arrayrulewidth}%
    \rule[-.05\ht\strutbox]{\result\maxlentime}{.95\ht\strutbox}}}%
\makebox[\dimexpr\maxlentime-0.2\tabcolsep+\arrayrulewidth][c]{#3}}

\newcommand\lsdtimebar[3][SeaGreen!30]{%
\FPeval\result{round((#3-0)/#2:4)}%
\rlap{\textcolor{#1}{\hspace*{\dimexpr-\tabcolsep+.5\arrayrulewidth}%
    \rule[-.05\ht\strutbox]{\result\maxlentime}{.95\ht\strutbox}}}%
\makebox[\dimexpr\maxlentime-0.2\tabcolsep+\arrayrulewidth][c]{#3}}

\def\headertime{New}
\AtBeginDocument{%
  }

\setcopyright{acmlicensed}
\acmJournal{IMWUT}
\acmYear{2025} \acmVolume{9} \acmNumber{2} \acmArticle{60} \acmMonth{6}\acmDOI{10.1145/3729462}

\acmISBN{978-1-4503-XXXX-X/2018/06}

\begin{document}

\pagenumbering{arabic}
\title{\sysname: Ultrasound-Enhanced Speech with Minimal Human Effort via Cross-Modal Synthesis}

\author{Luca Jiang-Tao Yu}
\authornote{Co-first author.}
\affiliation{
  \institution{The University of Hong Kong}
  \country{Hong Kong SAR, China}
  }
\email{lucayu@connect.hku.hk}
\orcid{0009-0004-3964-5874}

\author{Running Zhao}
\authornotemark[1]
\affiliation{
  \institution{The University of Hong Kong}
  \country{Hong Kong SAR, China}
  }
\email{rnzhao@connect.hku.hk}
\orcid{0000-0003-2496-3429}

\author{Sijie Ji}
\affiliation{
  \institution{The University of Hong Kong}
  \country{Hong Kong SAR, China}
  }
\email{sijieji@caltech.edu}
\orcid{0000-0002-6615-1982}

\author{Edith C.H. Ngai}
\affiliation{
  \institution{The University of Hong Kong}
  \country{Hong Kong SAR, China}
  }
\email{chngai@eee.hku.hk}
\orcid{0000-0002-3454-8731}

\author{Chenshu Wu}
\authornote{Corresponding author.}
\affiliation{
  \institution{The University of Hong Kong}
  \country{Hong Kong SAR, China}
}
\email{chenshu@cs.hku.hk}
\orcid{0000-0002-9700-4627}

\renewcommand{\shortauthors}{Yu and Zhao, et al.}

\begin{abstract}
    \rev{Speech enhancement is crucial for ubiquitous human-computer interaction. Recently, ultrasound-based acoustic sensing has emerged as an attractive choice for speech enhancement because of its superior ubiquity and performance. However, due to} inevitable interference from unexpected and unintended sources during audio-ultrasound data acquisition, existing solutions rely heavily on human effort for data collection and processing. This leads to significant data scarcity that limits the full potential of ultrasound-based speech enhancement. To address this, we propose \sysname, a cross-modal ultrasound synthesis framework for speech enhancement with minimal human effort. At its core is a two-stage framework that establishes the correspondence between visual and ultrasonic modalities by leveraging audio as a bridge. This approach overcomes challenges from the lack of paired video-ultrasound datasets and the inherent heterogeneity between video and ultrasound data. Our framework incorporates contrastive video-audio pre-training to project modalities into a shared semantic space and employs an audio-ultrasound encoder-decoder for ultrasound synthesis. We then present a speech enhancement network that enhances speech in the time-frequency domain and recovers the clean speech waveform via a neural vocoder. Comprehensive experiments show \sysname achieves remarkable performance using synthetic ultrasound data comparable to physical data, outperforming state-of-the-art ultrasound-based speech enhancement baselines. 
    \sysname is open-sourced at \href{https://github.com/aiot-lab/USpeech/}{\texttt{https://github.com/aiot-lab/USpeech/}}.
\end{abstract}



\received{20 February 2007}
\received[revised]{12 March 2009}
\received[accepted]{5 June 2009}

\maketitle

\section{Introduction}
\label{sec:intro}

Speech is the key enabler for human-computer interaction (HCI) with its potential to create more natural, accessible, and engaging interactions between humans and smart devices \rev{like mobile phones} \cite{munteanu2017designing, ephraim1995signal, xu2013experimental}. Speech-based mobile applications such as virtual assistants \cite{iannizzotto2018vision}, smart home control \cite{bajpai2019smart, fourniols2018overview}, and speech translation software \cite{wahlster2013verbmobil, nakamura2006atr} are revolutionizing our daily life. However, speech signals captured in real-world environments are often corrupted by various types of noise and interference, such as background noise, reverberation, and competing speakers, \rev{well-known as the cocktail party problem}. Unlike the human auditory system, it is challenging for machines to filter out the noise and irrelevant speech to pick up the desired speech command \cite{allen1994humans}. Consequently, speech enhancement, which aims to improve the quality and intelligibility of desired speech signals, \rev{becomes} an essential and critical processing step for many speech-related HCI applications \cite{sun2020supervised, martinez2022audio,chin2012audio}.

Decades of efforts have been devoted to audio-only solutions \cite{mmse, spectralsubtraction, kalman, yin2020phasen, TCNN, SEGAN, Convtasnet, sepformer}.
Recently, with the development of multimodal learning, combining other modalities reflecting vocal information demonstrates additional gains for audio-only solutions, such as video \cite{lipse3}, accelerometers \cite{bonesensorse3}, millimeter-wave (mmWave) radar \cite{radioses}, magnetic resonance imaging (MRI) and ultrasound tongue imaging (UTI\footnote{UTI is an imaging technique, employs ultrasound via specialized equipment \cite{SeeingspeechUTI}, different from the pseudo-ultrasound signals up to 24 kHz emitted by ubiquitous devices, the focus of \sysname. We will use the terms \textit{pseudo-ultrasound} and \textit{ultrasound} interchangeably in this paper.}) \cite{stone2005guide}. 
At a high level, these secondary modalities capture speech-induced motions at the desired \textit{source} where voice is generated, providing complementary contexts to audio recorded by a microphone at the \textit{destination} where all environmental sounds are mixed. 
Nonetheless, these modalities suffer from different drawbacks. Videos are privacy-intrusive and susceptible to lighting conditions. 
MRI, UTI, and mmWave radar rely on specialized equipment unavailable on ubiquitous mobile devices. Accelerometers are ubiquitous, but suffer from low sampling frequency and low sensitivity, preventing them from capturing fine-grained vibrations of speech.
\begin{figure}[t]
    \centering
    \includegraphics[width=1\linewidth]{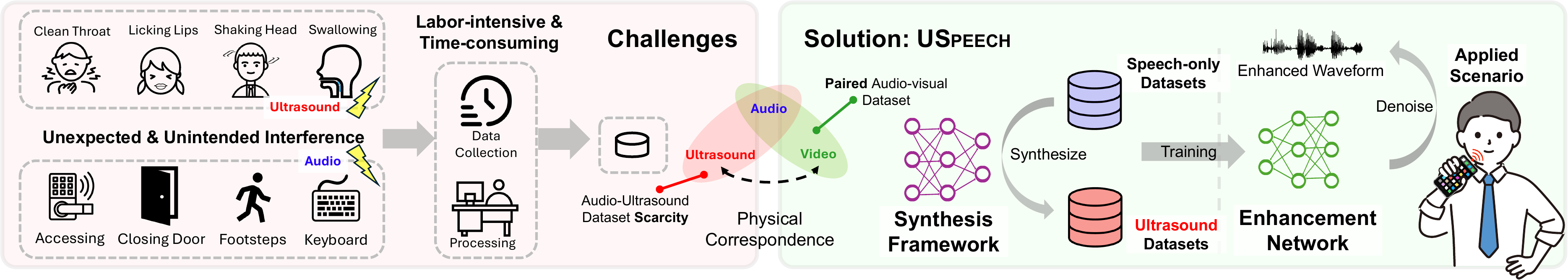}
    \vspace{-0.1in}
    \caption{An illustration of challenges in the audio-ultrasound dataset construction and an overview of \sysname design.}
    \label{fig:fig1}
\end{figure}

Differently, ultrasound-enhanced speech has emerged as a promising technology recently \cite{sun2021ultrase, ding2022ultraspeech}, capitalizing on the relationship between ultrasound and audio to improve the clarity and intelligibility of speech signals. 
Ultrasound-based solutions leverage acoustic sensing \cite{acousticsensingsurvey, acousticsensing} on commodity speakers and microphones without requiring any extra hardware, maintaining superior ubiquity while offering better performance. 
Specifically, the speaker emits a pseudo-ultrasound signal that is inaudible to the human ear, while the microphone captures both the audible speech and the reflected ultrasound signal. This signal encodes the information about the speaker's articulatory motions, which can then be used to further enhance the quality of the speech.

Existing ultrasound-based solutions \cite{sun2021ultrase, ding2022ultraspeech} rely heavily on human effort for \rev{manual} data collection and processing, which is both labor-intensive and time-consuming. This is due to the unavoidable occurrence of unexpected and unintended interferences during the collection of audio-ultrasound data (Fig. \ref{fig:fig1}). Unexpected interference, such as ambient noise from footsteps, closing the door, or even keyboard typing, can significantly degrade the quality of collected human speech. Additionally, unintended human actions like throat clearing, licking lips, \etc, can also introduce noise that may overpower the desired ultrasound signal. \rev{Moreover, the interfered data can} hinder the convergence speed of the network training and impair the performance \cite{xu2014regression, narayanan2014investigation}.
Therefore, once the interference occurs, researchers either need to control the interference factor precisely to re-collect the sample or put considerable effort into processing the data to mitigate the impact.
\rev{Consequently, building ultrasound datasets is both complex and prohibitively inefficient}, creating a notorious data scarcity issue in ultrasound-based speech enhancement.

In this paper, we aim to overcome this \rev{data scarcity problem and reduce human effort by building a synthesizer to generate reliable ultrasound data for speech enhancement}.
\rev{To achieve this, we identify videos as} an ideal lever for generating ultrasound, as they both share certain physical correspondence and represent articulatory gesture information -  video reflects the gesture in a visual format and ultrasound unveils that via Doppler shift.
However, the lack of paired video-ultrasound datasets poses a significant challenge for synthesizing ultrasound data from videos. Collecting such a dataset also requires extensive human effort.
Fortunately, a unique opportunity exists in large-scale paired video-audio datasets that are publicly available \rev{online}.
This motivates us to employ audio as the bridge to connect video and ultrasound. Specifically, we leverage the paired video-audio data with contrastive learning technology to inject the articulatory gestures from videos into audio embeddings, and then use a small set of audio-ultrasound paired data to activate the information. Such a framework maintains the physical correspondence between video and ultrasound while utilizing the inherent consistency between audio and ultrasound to reduce the heterogeneity between video and ultrasound data. Our key insight is that videos that capture articulatory gestures visually and ultrasounds that capture these gestures through Doppler shifts share a similar physical correspondence. Both modalities effectively represent articulatory gestures essential for speech enhancement.
Based on this, we propose a two-stage ultrasound synthesis framework that utilizes audio signals aligned with video signals in a semantic content space to generate ultrasound signals. \rev{In stage one, we employ contrastive video-audio pre-training with the proposed temporal-semantic dual loss to project the audio and video into a shared semantic content space while enhancing the temporal and semantic consistency.} Leveraging existing large-scale video-audio datasets, the first stage enables audio features to inherit the physical motion properties of video features. \rev{In stage two, we design an audio-ultrasound encoder-decoder network} to activate the physical motion information, constructing the connection between audio and ultrasound. \rev{Moreover, we propose a dual-MSE loss to enable stage two training to capture spectrogram static features and dynamic changes over time, generating ultrasound spectrograms that are accurate in both temporal and content levels.}

Building on top of the ultrasound synthesis framework, we advance to build an effective speech enhancement network using the generated ultrasound data.
We propose a speech enhancement network based on UNet architecture with Transformer layers to fuse the ultrasound and the noisy speech spectrogram and enhance the speech on the Time-Frequency (T-F) domain. %
Moreover, we apply the neural vocoder \cite{yamamoto2020parallel} to recover the speech waveform from the enhanced spectrogram directly.
Note that our enhancement network can be trained using the generated ultrasound data, reducing the human effort for physical data collection.

We present the complete implementation of \sysname, as shown in \fig \ref{fig:fig1}, which incorporates the ultrasound synthesis framework in tandem with the speech enhancement network. We prototype \sysname and evaluate it under various experiments. \sysname's speech enhancement network significantly outperforms state-of-the-art ultrasound-based speech enhancement baselines. At the same time, \sysname, when leveraging a synthetic dataset, can achieve performance \rev{comparable to} that obtained with a manually collected dataset. Furthermore, \sysname can boost speech enhancement on noisy large-scale speech-only datasets by leveraging the ultrasound data generated from them. 
Additionally, we provide various real-world examples at \href{https://aiot-lab.github.io/USpeech/}{\texttt{https://aiot-lab.github.io/USpeech/}}.
Our core contributions are summarized as follows:

\begin{itemize}
    \item We propose \sysname, the first cross-modal ultrasound synthesis framework for speech enhancement with minimal human effort of data collection and processing. %
    \item We design a two-stage ultrasound synthesis framework, which utilizes audio as a bridge to relate video and ultrasound for ultrasound synthesis. \rev{In stage one, we employ the contrastive video-audio pertaining with the proposed temporal-semantic dual loss to inject the articulatory gesture information from videos into audio embeddings, and in stage two, we design the audio-ultrasound encoder-decoder network with dual-MSE loss to activate the information for ultrasound synthesis.}
    Based on this, we introduce an effective speech enhancement network by fusing audio and ultrasound inputs, which can be trained from synthesized ultrasound data.

    \item \sysname is comprehensively evaluated in real-world scenarios under various settings. Experimental results show that \sysname outperforms state-of-the-art ultrasound-based speech enhancement baselines, while the performance on synthetic data is on par with that on physically collected data.
\end{itemize}

\section{Preliminaries}
\label{sec:preliminaries}

\begin{figure}[h]
    \centering
    \includegraphics[width=0.9\linewidth]{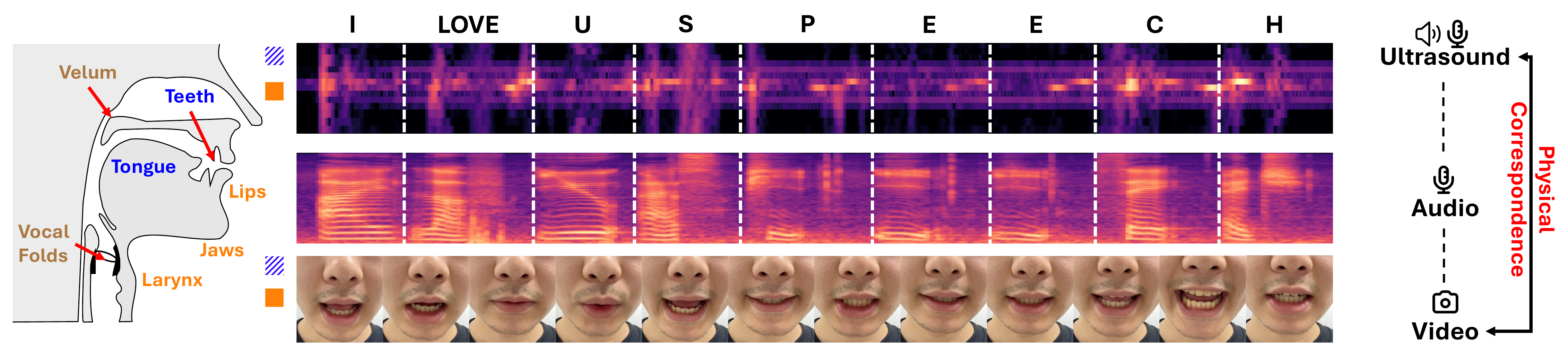}
    \caption{The vocal tract and the physical relationship among three modalities. \textcolor{orange}{\textbf{Orange}} indicates organs \textit{consistently visible}; \textcolor{blue}{\textbf{Blue}} indicates organs that are \textit{occasionally visible}; \textcolor{brown}{\textbf{Brown}} indicates \textit{invisible} organs.}
    \label{fig:vocal_tract}
\end{figure}

\subsection{Articulatory Gestures}
\rev{Articulatory gestures are the movements of vocal organs that produce phonemes, the fundamental sound units of speech. These gestures involve various articulators, such as the lips, teeth, jaw, velum, tongue, vocal folds, and larynx (\fig \ref{fig:vocal_tract}) \cite{gick2013articulatory}. Speech production can be simplified into two steps: the source generates an initial sound, and the vocal tract modulates it \cite{wolfe2020mechanics}. Different articulators create distinct phonemes, such as bilabial sounds (/p/, /b/, /m/) produced by lips, and sounds like /f/ and /v/ created by air passing between the bottom lip and upper teeth. The jaw and tongue collaborate to produce both consonants (/k/, /g/) and vowels (/i/, /u/). Articulatory gestures differ by sensing modality. Videos capture motions of visible organs, such as the jaws, larynx, and lips, with occasional visibility of the tongue and teeth (\fig \ref{fig:vocal_tract}) \cite{brooke1983analysis}. In contrast, ultrasound-based acoustic sensing captures fine-grained surface vibrations at high sampling rates, providing richer contexts for speech enhancement \cite{sun2021ultrase}.}

\begin{figure}[t]
    \subfloat[\mrev{Different unexpected and unintended interference on the audio and ultrasound spectrograms when collecting the audio-ultrasound dataset.}]{
    \label{fig:difficulties}
    \includegraphics[width=0.4\linewidth]{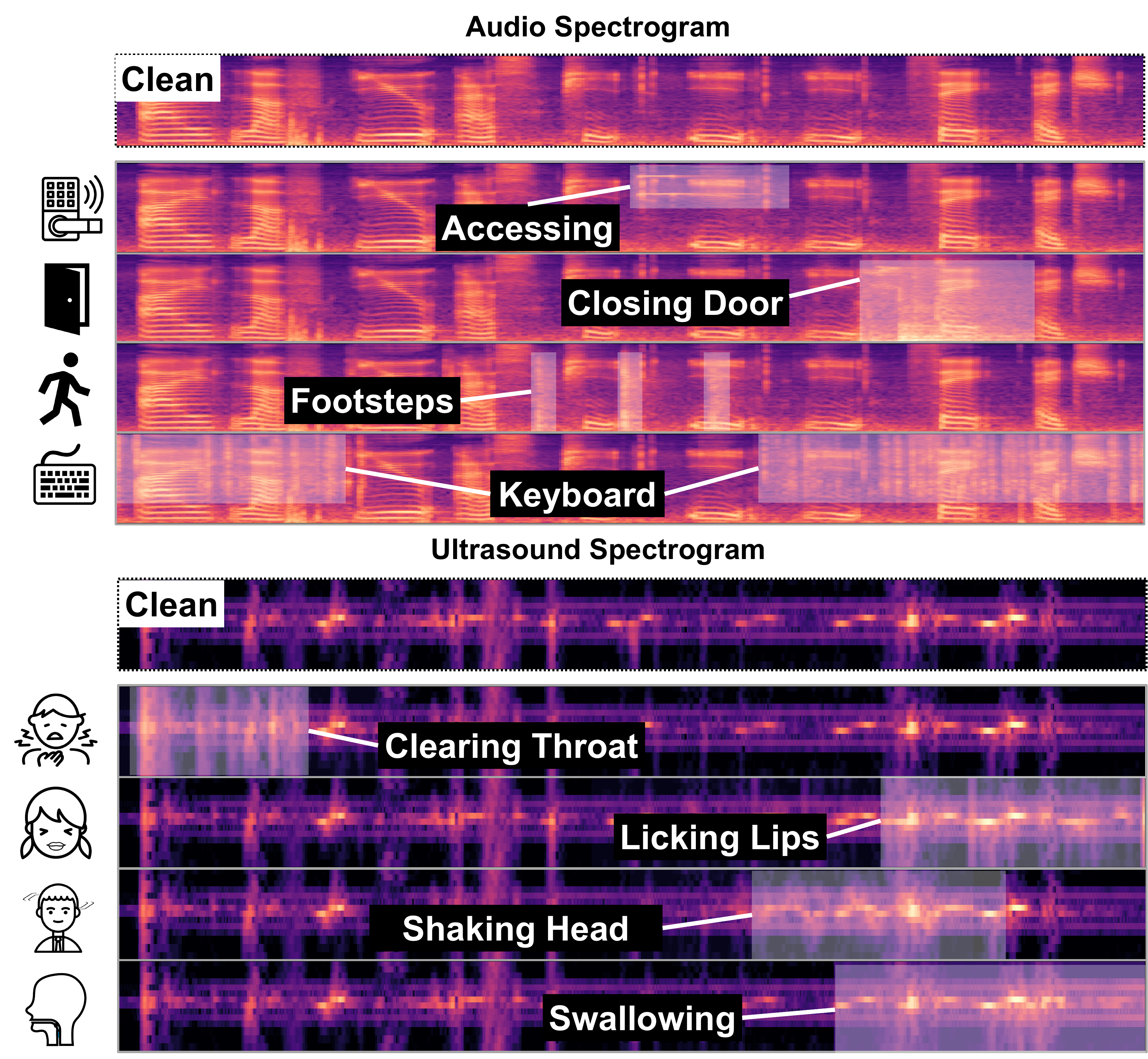}}
   \hspace{+0.3in}
    \subfloat[The test loss and results of model trained on the clean dataset and interfered dataset.]{
    \label{fig:prelinimary}
      \includegraphics[width=0.4\linewidth]{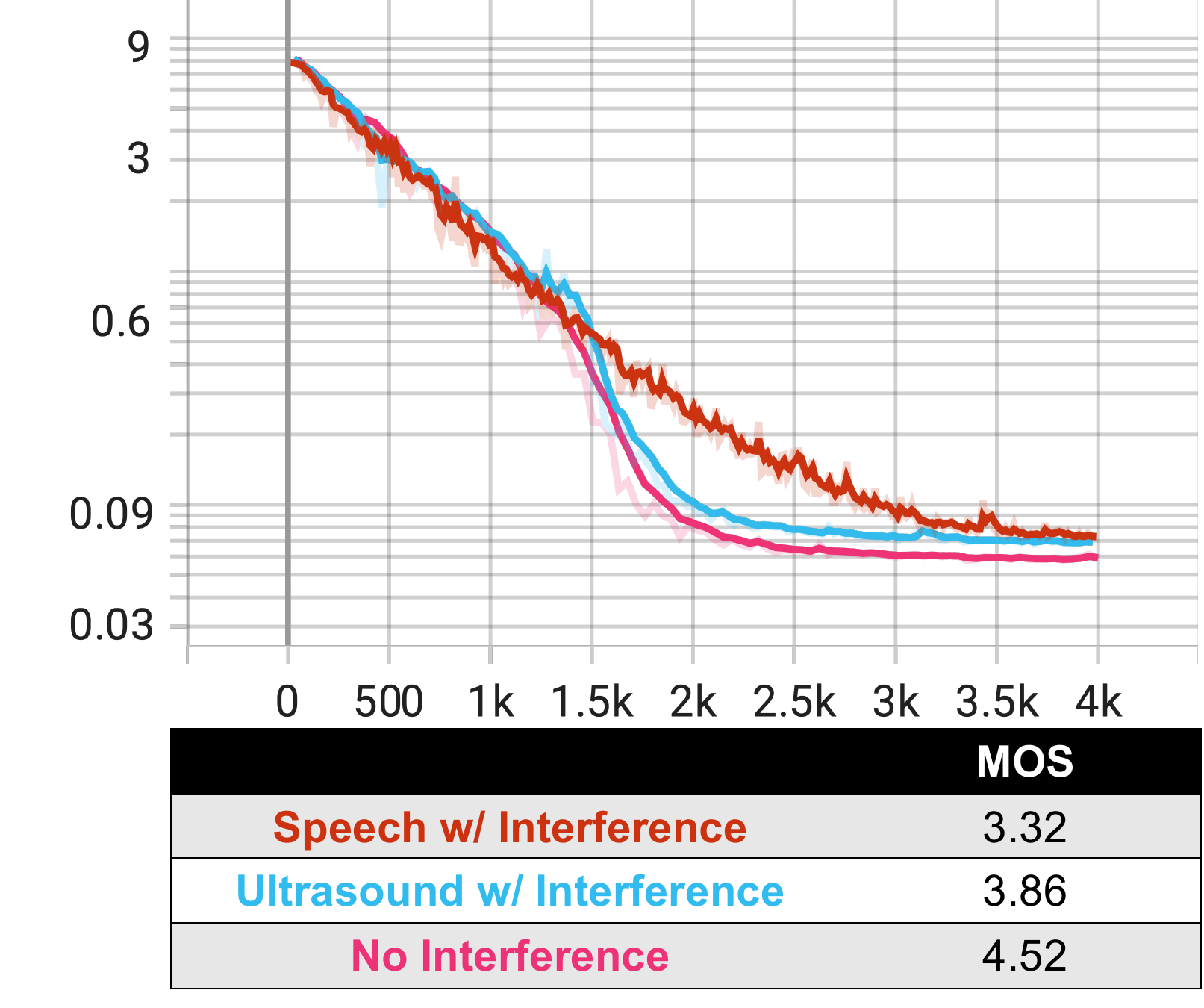}}

    \caption{Challenges and opportunities.}
    \vspace{-0.2in}
\label{fig:ch_and_op}
\end{figure}

\subsection{Ultrasound Articulatory Sensing}
\label{sec:ultrasound_articulatory_sensing}
The commodity speaker and microphone in mobile phones transmit the ultrasound wave and receive the reflected ultrasound signal modulated by articulatory gestures for further processing, respectively.
The articulatory gestures exhibit velocities ranging from -160 \textasciitilde 160 cm/s for bi-directional propagation \cite{teplansky2019tongue}. According to the Doppler effect, the velocity introduces about -94 \textasciitilde 94 Hz Doppler shift within the 20 kHz band. Various sensing waveforms like FMCW \cite{mao2016cat}, OFDM \cite{nandakumar2016fingerio}, and PN sequences \cite{yun2017strata, sun2018vskin} have demonstrated the ability to capture the impulse response but they all suffer from the low sampling rate to capture rapid changes in motion. 
Ultrasound Continuous Wave (CW) modulated Global System for Mobile Communications (GSM) sequence is applied to calculate the Channel Impulse Response \cite{ding2022ultraspeech}, which is constrained by its single-measurement approach that could lead to measurement errors. 
To address these challenges, particularly the issue of low sampling rates and the need for multiple measurements, we propose to apply the multi-tone CW, treating each tone as one independent measurement. \rev{The transmitting signal $s(t)$ and the received signal $r(t)$ can be formulated as follows:
$s(t) = \sum^{N-1}_{i=0} \cos \left[ 2 \pi (f_0 + i \Delta f) + \phi_i \right], r(t) = \sum^{N-1}_{i=0} \cos \left[ 2 \pi (1 + \frac{\Delta v}{c}) (f_0 + i \Delta f) + \phi_i + \Delta \phi_i \right],$
where the original frequency $f_0$ is set to 17.25 kHz, the frequency interval $\Delta f$ is set to 750 Hz, the number of tones $N$ is set to 8, and $\Delta v$ and $c$ indicate the velocity of articulatory gestures and sound speed. We set the sampling rate as 48 kHz. 
We then perform STFT to obtain the Doppler shift related to articulatory gestures.
}

\section{Challenges and Opportunities}
\label{sec:ch_and_op}

\subsection{Speech in Video, Audio, and Ultrasound}
\label{sec:speech_in_video_audio_and_ultrasound}
As mentioned above, both video and ultrasound modalities are capable of capturing articulatory gestures during speech production, with \fig \ref{fig:vocal_tract} illustrating the physical relationships between video and ultrasound. Taking the phoneme /v/ from the word \textit{"love"} as an example, we observe that its production involves high-frequency, fine-grained vibrations of the larynx coupled with lip closure. The specific movements correspond to the second interval on ultrasound recordings and are visually discernible in the second and third frames of the video. The observation validates the physical correspondence between video and ultrasound. 
However, paired video-ultrasound datasets with speech labels are missing, and more importantly, very costly to build. 
The different representation forms of video and ultrasound also pose a challenge for the modality transformation. 
For the same example, it can be seen in \fig \ref{fig:vocal_tract} that the audio spectrogram captures the distinct sound of the /v/ phoneme at a similar position. This reveals that audio shares similar representation forms of time series with ultrasound. 
This promises an opportunity to leverage audio as the bridge to relate video and ultrasound, thereby generating massive audio-ultrasound data from large-scale audio-video data, which are widely available online. By doing so, we can overcome all the drawbacks shown in \fig \ref{fig:fig1}.

\subsection{Difficulties in Ultrasound Dataset Construction}
\label{sec:difficulties_in_data_construction}

\mrev{Training on clean audio-ultrasound datasets is critical for achieving stable convergence and high perceptual quality. In a preliminary study, we examine how different types of interference affect model performance. We separately train three enhancement models using the same architecture but different datasets: (1) fully clean audio-ultrasound data, (2) a dataset with 50\% clean and 50\% speech-interfered data, and (3) a dataset with 50\% clean and 50\% ultrasound-interfered data. Each model is tested on a corresponding evaluation set that matches its interference type, except for the clean-trained model, which is tested on fully clean data. As shown in \fig \ref{fig:prelinimary}, the model trained and tested on clean data exhibits smoother and faster convergence in the loss function, reaching a lower final value. The Mean Opinion Score (MOS) also reflects this trend: the clean-trained model achieves a score of 4.52, while speech and ultrasound interference settings yield lower scores of 3.32 and 3.86, respectively. These results suggest that clean training data is beneficial for learning stable and high-quality enhancement models.
However, collecting such clean data in practice is highly challenging. Audio and ultrasound spectrograms are extremely sensitive to environmental and physiological factors, such as background noise (\eg, footsteps, keyboard typing) and inadvertent human actions (\eg, throat clearing, lip licking). As shown in \fig \ref{fig:difficulties}, these interferences often overwhelm the target signal, degrading data quality and complicating model training. Moreover, they increase the burden of manual processing and annotation, ultimately limiting the scale and quality of the collected dataset. This issue is reflected in the performance drops observed in \tab \ref{tab:scale_en}, where smaller and noisier datasets lead to degraded metrics.
Although such interferences are inevitable in real-world applications, training on clean data remains beneficial for building reliable models. To enable training on clean data without significant data collection efforts, we propose a two-stage ultrasound synthesis framework to reduce reliance on physically collected clean ultrasound. We leverage the articulatory correspondence between video and ultrasound and use audio as a bridge to synthesize ultrasound representations. A detailed example of the difficulty in constructing a high-quality video-ultrasound dataset is shown in \fig \ref{fig:ch_video_ultrasound} in the Appendix.
}

\begin{figure}[t]
    \centering
    \includegraphics[width=0.85\linewidth]{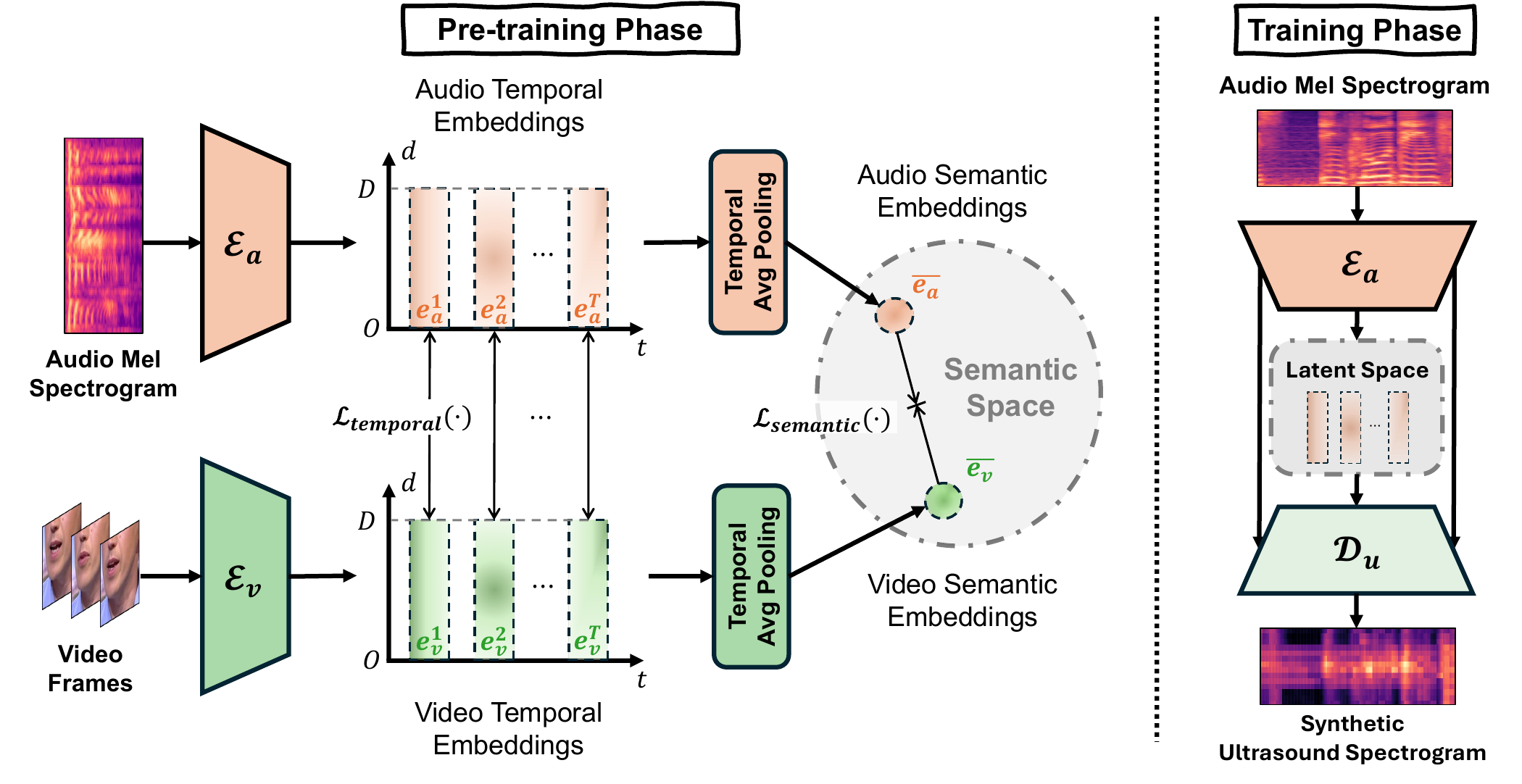}
    \caption{An overview of ultrasound synthesis framework. It includes the contrastive video-audio pre-training phase (left) for projecting video and audio into a shared semantic content space, and the training phase (right) about the audio-ultrasound encoder-decoder network for constructing the connection between audio and ultrasound.}
    \label{fig:synthesis}
\end{figure}

\section{\sysname Overview}
\label{sec:overview}

In this section, we elaborate on the overall structure of \sysname, which contains an ultrasound synthesis framework (\S\ref{sec:ultrasound_synthesis}) and a speech enhancement network (\S\ref{sec:speech_enhancement}).

\head{Ultrasound Synthesis Framework} 
The synthesis framework contains two stages (see \fig \ref{fig:synthesis}):

\noindent\textit{1) Contrastive Video-Audio Pre-training:} 
The contrastive video-audio pertaining projects the audio and video into a shared semantic content space. 
It utilizes open-source video-audio datasets containing the visual articulatory gestures to pre-train the video encoder and audio encoder via contrastive learning. 
This learning process injects the articulatory gestures from videos into audio embeddings, enabling the audio encoder to inherit the physical motion properties.

\noindent\textit{2) Audio-Ultrasound Encoder-Decoder Network:} After contrastive video-audio pre-training, the pre-trained audio encoder is capable of encoding the audio into embeddings containing physical properties of articulatory gestures. Subsequently, to adapt the information from audio to ultrasound, we design a UNet-based network with the pre-trained audio encoder and the ultrasound decoder to transform the above embeddings into the ultrasound Doppler shift spectrogram.

\head{Speech Enhancement Network}
The enhancement network contains two phases (see \fig \ref{fig:enhancement}):

\noindent\textit{1) Mel Spectrogram Enhancement:} 
The first phase is a UNet network that enhances the Mel spectrogram by taking ultrasound and noisy speech as input. The encoded embeddings of ultrasound and noisy speech are fused by the Transformer layers and then, the fused embeddings are decoded into Mel spectrogram of enhanced speech.

\noindent\textit{2) Waveform Recovery:} In the second phase, we leverage a neural vocoder to recover the high-quality enhanced speech waveform from the enhanced Mel spectrogram.

\section{\sysname Design}
\label{sec:design}
In this section, we commence by detailing the synthesis process of ultrasound. Following this, we detail the design of the speech enhancement network. \rev{We provide the model details in Appendix \S \ref{sec:appendix_model_details}.}

\subsection{Ultrasound Synthesis}
\label{sec:ultrasound_synthesis}

\subsubsection{Contrastive Video-Audio Pre-training}
\label{sec:contrastive_pretraining}
\rev{
Due to the independent training process, existing video or image and audio backbones struggle to reflect the alignment between video and audio modalities, leading to a lack of shared representation space \cite{wang2020alignnet, gong2022uavm}. To address this, we propose a contrastive learning framework that enforces temporal and semantic consistency, ensuring that video embeddings capture articulatory gestures while aligning them with audio representations. Our method first extracts articulatory features using a dual-branch network.} For the video encoder, inspired by SlowOnly \cite{feichtenhofer2019slowfast}, we map mouth-region frames ($128 \times 128$) into a 512-dimensional embedding. The model consists of five 3D convolutional layers followed by batch normalization and ReLU activations, gradually reducing spatial dimensions while preserving temporal correlations.
For the audio encoder, adapted from PANNs \cite{kong2020panns}, originally pre-trained on the Audioset dataset \cite{gemmeke2017audio}, it employs six 2D convolutional blocks with batch normalization, ReLU activations, and average pooling. The extracted features are refined through both average and max pooling operations, ensuring no loss of subtle articulatory information. The pooled outputs are further processed by two fully connected layers (2048 and 512 dimensions), enhancing representation power before alignment with video embeddings.
By integrating these components, our framework effectively aligns video and audio modalities, facilitating a shared representation space that captures the nuanced interplay between visual articulatory gestures and corresponding audio signals.

\rev{Given the independent training of video and audio backbones, direct alignment leads to suboptimal performance due to the lack of fine-grained articulatory synchronization. Without temporal-semantic constraints, embeddings from different modalities may not be sufficiently aligned, leading to misrepresentation in ultrasound synthesis. This misalignment results in a degraded performance in downstream tasks. By introducing the temporal-semantic dual loss, we improve cross-modal coherence, ensuring that articulatory gestures in video are properly synchronized well with corresponding phonemes in audio.} The dual loss consists of a temporal loss $\mathcal{L}_{temporal}$, which aligns fine-grained articulatory dynamics at each frame, reducing misalignment errors, and a semantic loss $\mathcal{L}_{semantic}$, which enforces similarity between global video-audio embeddings while contrasting unrelated pairs.
Both losses are based on the InfoNCE formulation \cite{oord2018representation}:
\begin{equation}
\label{eq:temporal}
\mathcal{L}_{temporal}^{(i,j)} = -\frac{1}{2} \log \frac{\exp(\text{sim}(e_a^i, e_v^j) / \tau)}{\sum_{k=1}^{n_t}\exp(\text{sim}(e_a^i, e_v^k) / \tau)} - \frac{1}{2} \log \frac{\exp(\text{sim}(e_a^i, e_v^j) / \tau)}{\sum_{k=1}^{n_t}\exp(\text{sim}(e_a^k, e_v^j) / \tau)},
\end{equation}

\begin{equation}
\label{eq:semantic}
\mathcal{L}_{semantic}^{(i,j)} = -\frac{1}{2} \log \frac{\exp(\text{sim}(\overline{e_a}^i, \overline{e_v}^j) / \tau)}{\sum_{k=1}^{n_s}\exp(\text{sim}(\overline{e_a}^i, \overline{e_v}^k) / \tau)} - \frac{1}{2} \log \frac{\exp(\text{sim}(\overline{e_a}^i, \overline{e_v}^j) / \tau)}{\sum_{k=1}^{n_s}\exp(\text{sim}(\overline{e_a}^k, \overline{e_v}^j) / \tau)},
\end{equation}
where $\text{sim}(\cdot)$ is cosine similarity, $n_t$ and $n_s$ are temporal and semantic segment counts, and $\tau$ (default 0.07) is the temperature parameter.
The final contrastive loss combines both terms:
\begin{equation}
\label{eq:contrastive_loss}
\mathcal{L}_{contrastive} = \lambda \mathcal{L}_{temporal} + (1 - \lambda) \mathcal{L}_{semantic},
\end{equation}
where $\lambda$ (default 0.5) balances temporal alignment and semantic consistency.

\begin{figure*}[t]
    \centering
    \includegraphics[width=0.75\linewidth]{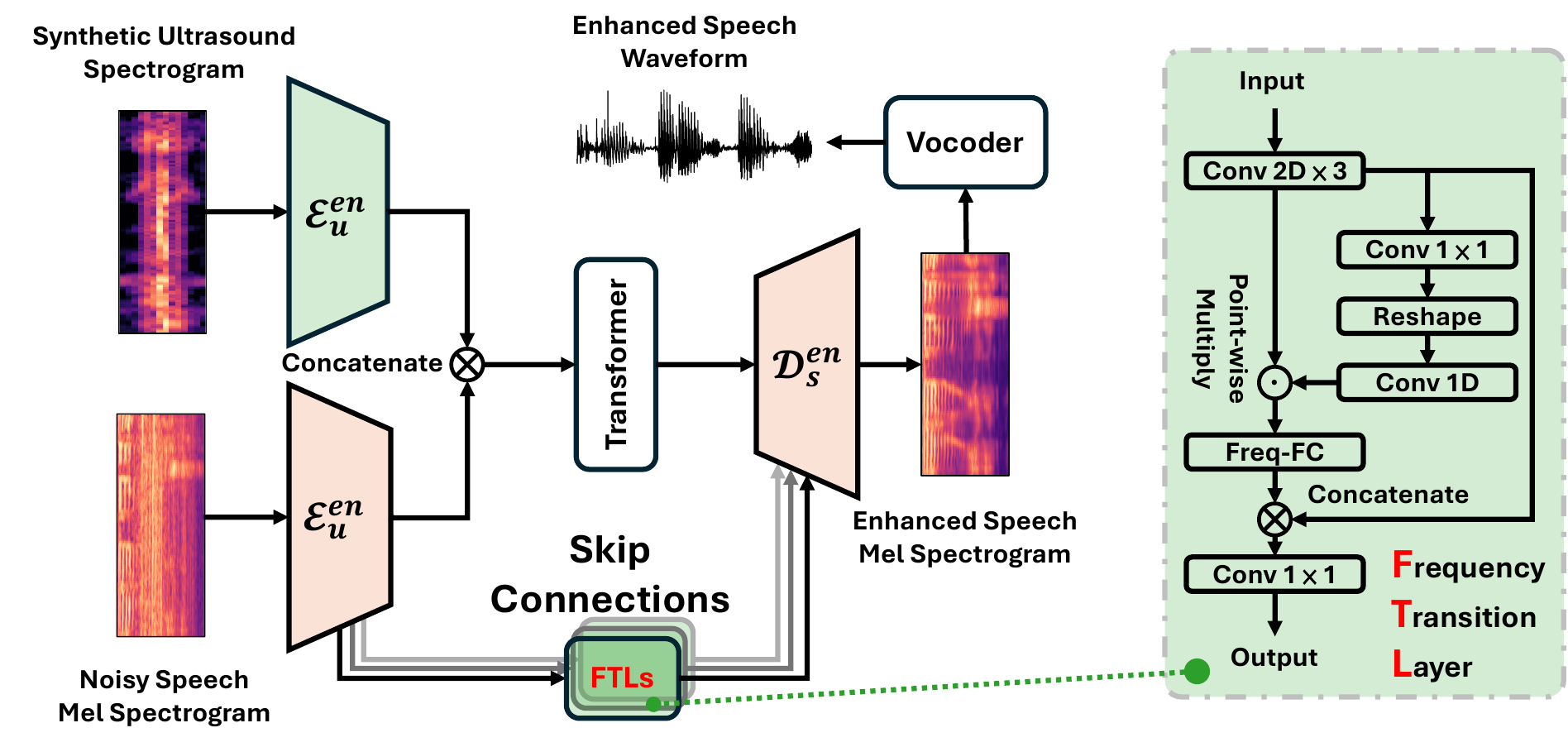}
    \caption{An overview of speech enhancement network, including Mel spectrogram enhancement and waveform recovery (left). The architecture of the Frequency Transition Layer (FTL) is shown on the right.}
    \label{fig:enhancement}
\end{figure*}

\subsubsection{Audio-Ultrasound Encoder-Decoder Network.}
\label{sec:audio_ultrasound_encoder_decoder_network}
\rev{After contrastive learning aligns video-audio representations temporally and semantically, the pre-trained audio encoder effectively captures articulatory information from Mel spectrograms. This alignment enables the direct deployment of the encoder for synthesizing ultrasound Doppler shift spectrograms, which is particularly beneficial when working with a small-scale dataset.} Speech and ultrasound signals occupy distinct frequency bands, so we apply an 8th-order elliptic filter (lowpass: 8 kHz, highpass: 16 kHz) to preserve articulatory motion in the ultrasound data. The speech signal is downsampled from 48 kHz to 16 kHz and transformed into Mel spectrograms ($x_s \in \mathbb{R}^{T \times 128}$). Ultrasound signals undergo STFT (FFT: 4096, window: 4080, hop: 240) with a frequency resolution of 11.72 Hz. We remove the central and adjacent bins ($\pm 11.72, 0$) Hz to suppress static reflections, retaining $7 \times 2$ bins per tone for articulatory gestures ($x_u^T \in \mathbb{R}^{T \times 14}$). The ultrasound decoder $\mathcal{D}_u$ is trained with the pre-trained speech encoder $\mathcal{E}_a$ using $(x_s, x_u)$ pairs.

\rev{
The use of contrastive pre-training can transfer knowledge from a large-scale audio-video dataset, which aligns the temporal and semantic features of both modalities. This significantly enhances the robustness of the encoder, even with limited ultrasound data.} Moreover, we use the UNet architecture \cite{ronneberger2015u} to maintain spatial consistency and improve feature retention during decoding. The skip connections are employed to concatenate feature maps from the corresponding encoder and decoder layers. This allows fine-grained preservation of temporal and dynamic articulatory features, essential for high-quality ultrasound synthesis. The decoder, adopting a transposed convolutional architecture with six layers and a final projector (kernel: $2 \times 1$, stride: $2 \times 1$, padding: 25, dilation: 1), refines the ultrasound spectrogram resolution, ensuring accurate articulation representation and reducing blurring effects during spectrogram reconstruction.
\rev{The synthesized ultrasound spectrogram $d_{\text{syn}} \in \mathbb{R}^{T \times F}$ must effectively capture micro-scale temporal variations essential for articulation tracking. Traditional MSE loss struggles with this, as it primarily minimizes absolute differences between predicted and ground-truth values, neglecting the rapid transitions of articulatory motion, especially for the ultrasound Doppler frequency shift spectrogram that reflects the subtle and sensitive articulatory. This limitation often results in over-smoothing, causing a loss of fine-grained speech dynamics.} To address this, we introduce a dual-MSE loss function that optimizes both the spectrogram and its temporal derivative:
\begin{equation}
    \mathcal{L}_{\text{dual-MSE}} = \alpha \cdot \text{MSE}(d_{\text{syn}}, d_{\text{gt}}) + (1 - \alpha) \cdot \text{MSE}(\Delta d_{\text{syn}}, \Delta d_{\text{gt}}),
\end{equation}
where $\alpha$ (default 0.5) balances spectral accuracy and preservation of fine temporal structures. \rev{Unlike standard MSE loss, incorporating $\Delta d_{\text{syn}}$ ensures that the model maintains critical articulatory variations, preventing the degradation of speech intelligibility due to excessive smoothing. Without this term, the model fails to reconstruct rapid transitions, leading to blurring effects that obscure essential articulation details.
This approach significantly improves synthesis fidelity, demonstrating that preserving micro-scale temporal variations enhances spectral resolution and articulation clarity. Consequently, \sysname achieves ultrasound spectrogram synthesis quality comparable to physically collected data.
}

\subsection{Speech Enhancement}
\label{sec:speech_enhancement}

\subsubsection{Mel Spectrogram Enhancement}
\label{sec:t_f_mel_spectrogram_enhancement}
Based on the UNet architecture, we have developed a network for enhancing the Mel spectrogram, as illustrated in \fig \ref{fig:enhancement}. Recognizing the critical role of harmonic correlation across the frequency axis in augmenting the quality of the T-F spectrogram \cite{wakabayashi2018single}, our design incorporates both a UNet-based backbone for contextual temporal information capture and Frequency Transition Layers (FTL) \cite{zhao2022radio2speech}. The FTLs are strategically positioned between each downsampling convolutional layer to enhance harmonic information capture. The FTL consists of three stacked convolutional layers, a fully connected layer used as a transformation matrix, and a $1 \times 1$ convolutional layer as the output, shown in \fig \ref{fig:enhancement}. Each convolutional layer is comprised of a 2D convolutional layer, followed by batch normalization and a ReLU activation function. Parallel to the speech encoder branch, the ultrasound branch mirrors these settings without FTLs. A key feature of our design is the integration of the Vision Transformer \cite{vaswani2017attention, dosovitskiy2020image} at the bottleneck. This introduces an attention mechanism to the temporal segments of the feature map, leveraging the Transformer's capabilities pre-trained on ImageNet \cite{deng2009imagenet}. The enhancement of the Mel spectrogram is further refined by applying MSE loss.

\begin{figure}[t]
    \centering
    \includegraphics[width=0.5\linewidth]{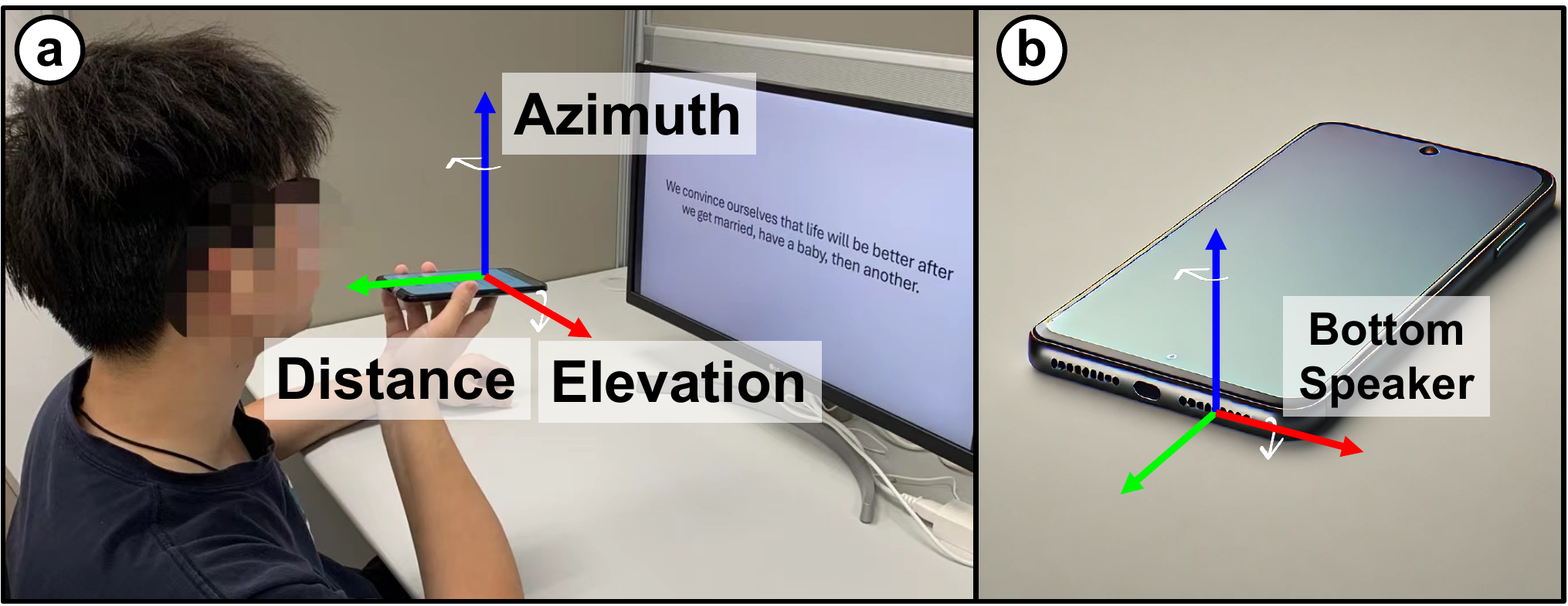}
    \caption{Data collection scenarios. Three axes define the device's coordinate system. The axis depicted in \textcolor{green}{\textbf{Green}} denotes the distance to the mouth. Rotations about the \textcolor{red}{\textbf{Red}} axis represent elevation adjustments, and rotations around the \textcolor{blue}{\textbf{Blue}} axis correspond to changes in azimuth.}
    \label{fig:collection}
    \vspace{-0.1in}
\end{figure}

\subsubsection{Waveform Recovery}
\label{sec:waveform_recovery}
After enhancing the T-F domain spectrogram, we employ a neural vocoder, specifically the Parallel WaveGAN \cite{yamamoto2020parallel}, to reconstruct the waveform from the enhanced Mel spectrogram. The phase information plays a pivotal role in accurately recovering waveform from T-F domain spectrograms \cite{paliwal2011importance}. However, converting audio signals into Mel spectrograms inherently results in the loss of phase information, which is crucial for reconstructing high-quality natural-sounding waveforms. To overcome this challenge, we employ a neural vocoder that includes both a generator and a discriminator. The generator synthesizes the audio waveform directly from the Mel spectrogram by learning the mapping between the spectrogram and time-domain waveform, while the discriminator helps ensure the generated waveforms sound realistic by distinguishing between real and generated audio. By leveraging adversarial training, the neural vocoder learns not only to reconstruct amplitude but also implicitly recovers the missing phase information. This allows for more accurate and perceptually natural waveform generation, compensating for the phase loss that occurs during the spectrogram conversion process. The vocoder undergoes training through the joint optimization of a multi-resolution spectrogram loss and an adversarial loss \cite{yamamoto2020parallel}. This dual loss approach significantly enhances the model’s ability to replicate the time-frequency distribution characteristic of realistic speech waveforms.

\section{Experimental Setup}
\label{sec:experimental_setup}

\subsection{Data Collection}
\label{sec:data_collection}

Due to the lossy M4A compression method used by the recorded in standard smartphones, which is not capable of capturing the ultrasound band, we develop an Android application to gather an audio-ultrasound dataset using the OPPO Reno2 Z. \rev{Given that a typical smartphone's earpiece power (about 5 $\mu$W) is too low for sensing purposes, we instead utilized the more powerful bottom-mounted speaker (1 W) along with the microphone for transmitting and receiving signals.} This setup enables us to record the synchrony between ultrasound articulatory and speech accurately. The data collection setup is illustrated in \fig \ref{fig:collection}.

We recruit 7 volunteers (2 females and 5 males) to participate in the data collection by reading the selected materials, aiming to construct an audio-ultrasound dataset in an environment with about 50 dB of ambient noise. The experimental setting is, regardless of the azimuth and elevation, within a reasonable range of 5 to 10 centimeters between the mouth and the bottom microphone, with no device rotation. The reading materials comprised a mix of articles, news, and other reading materials, outlined in the Appendix \tab \ref{tab:materials}. All volunteers read these materials in the same order, and the volume of the dataset varies from 70 to 83 dB. Overall, we collected 2.4 hours of audio-ultrasound data for the ultrasound synthesis framework (\S\ref{sec:ultsound_synthesis_dataset}) and 2.4 hours of audio-ultrasound data for the speech enhancement network (\S\ref{sec:speech_enhancement_se_dataset}), which are different materials to prevent data leakage and overlaps.

\subsection{Data Preparation}
\label{sec:data_preparation}

\subsubsection{\mrev{Audio-Video Pre-training Dataset}} The LRW dataset \cite{Chung16} is the public video-audio dataset that is utilized in the contrastive video-audio pre-training phase (\S \ref{sec:contrastive_pretraining}), which comprises approximately 1,000 utterances of 500 distinct words, each with a fixed duration of 1.22 seconds audio and 29 video frames.

\subsubsection{Ultrasound Synthesis Dataset}
\label{sec:ultsound_synthesis_dataset} 
We use the collected 2.4-hour audio-ultrasound dataset as the ultrasound synthesis dataset. We adopt a \mrev{temporal-based 80\%-20\% split for each volunteer, where the first 20\% of each participant’s recordings are used as the test set and the remaining 80\% as the training set for the audio-ultrasound encoder-decoder network (\S\ref{sec:audio_ultrasound_encoder_decoder_network})}. We further ensure that the speech materials used in the training and testing sets are entirely \textbf{\textit{non-overlapping}} to prevent data leakage.

\subsubsection{Speech Enhancement (SE) Dataset}
\label{sec:speech_enhancement_se_dataset}
\rev{To assess the enhancement performance of \sysname, we employ multiple datasets. First, we gather the physical SE dataset to train the speech enhancement network and evaluate the effectiveness of the proposed \sysname. Second, we utilize the ultrasound synthesis framework, trained on the ultrasound synthesis dataset in \S \ref{sec:ultsound_synthesis_dataset}, to generate synthetic ultrasound spectrograms from the physical SE dataset, thereby creating the synthetic SE dataset. Finally, we apply the same procedure to large-scale speech-only datasets to generate corresponding ultrasound spectrograms, constructing large-scale synthetic SE datasets.}
\label{sec:handicraft_noisy_dataset}
\begin{itemize}
    \item \textbf{Physical SE Dataset}: \rev{We use the collected 2.4-hour audio-ultrasound dataset (different from the Ultrasound Synthesis Dataset) as the physical SE dataset, where 80\% (1.92 hours) to train the speech enhancement network (\S\ref{sec:speech_enhancement_se_dataset}) and 20\% (0.48 hours) to test, \mrev{which split via teamporal-based as well.}} We refer to this dataset as "physical" or "phy." in experimental evaluation because it utilizes real-world collected ultrasound spectrograms. The dataset does not overlap with the ultrasound synthesis dataset to prevent data leakage. To adapt the input of other baselines, we also make the complex physical SE dataset with the same contents while keeping the imaginary part of spectrograms.
    
    \item \textbf{Synthetic SE Dataset}: Leveraging the ultrasound synthesis framework trained on the ultrasound synthesis dataset, we generate the synthetic SE dataset from the physical SE dataset, where the collected audio Mel spectrograms serve as input and the corresponding ultrasound spectrograms are generated as output. We then pair the audio Mel spectrograms with the synthetic ultrasound spectrograms to form the synthetic SE dataset. \rev{To ensure the effectiveness of evaluation, the training set of the synthetic SE dataset mirrors the training set of the physical SE dataset, with the ultrasound spectrograms from the physical data replaced by synthetic spectrograms. The test set, however, remains identical to that of the physical SE dataset.}

    \item \textbf{Large-scale Synthetic SE Datasets}: We employ three distinctive, public, large-scale speech-only datasets for ultrasound generation to evaluate the generalizability of the ultrasound synthesis framework in \S \ref{sec:large-scale_dataset}. Similar to the synthetic SE dataset, we use the trained synthesis model to generate ultrasound spectrograms from these speech-only datasets and pair the audio and ultrasound to construct the large-scale synthetic SE datasets. \rev{To maintain the effectiveness of evaluation, the entire large-scale synthetic SE datasets are used as the training set, while the test set of the physical SE dataset is exclusively used for testing.} The overview of the training sets of the large-scale speech-only datasets is as follows:
    \begin{itemize}
        \item \textbf{LJSpeech} \cite{ljspeech17} (24 hours) consists of 13.1k short audio clips from a single female speaker reading passages from 7 non-fiction books. This dataset provides sufficient scale but lacks speaker diversity.
    
        \item \textbf{TIMIT} \cite{garofolo1993timit} (5.8 hours) addresses this limitation by including recordings from 630 speakers (approximately 70\% male and 30\% female) with diverse features such as gender, dialect, and age. Each speaker has 10 utterances, offering high speaker diversity but on a smaller scale than LJSpeech.
    
        \item \textbf{VCTK} \cite{yamagishi2012} (44 hours) contains speech data from 110 speakers with various accents, each reading approximately 400 sentences derived from news articles and other reading materials. This dataset offers both substantial scale and speaker diversity. 
\end{itemize}

\end{itemize}

For speech enhancement network training, we generate the \textbf{Noisy SE Dataset} by mixing the clean speech we collected with the noise source dataset Nonspeech7k \cite{rashid2023nonspeech7k}. The dataset contains \textit{human's nonspeech}, \eg screaming, crying, coughing, from freesound.org, YouTube, and Aigei, rather than the ambient noise, which has a more similar distribution with the speech. Each clean speech sample is mixed with 20 different noises in collected physical and synthetic SE datasets and 1 noise in large-scale synthetic SE datasets, and the mixing process is with the specific SNRs randomly chosen from $\{-10, -5, 0, 5, 10, 15\}$ dB.

\subsection{Baselines}
\label{sec:baselines}

To illustrate the effectiveness of the proposed \sysname, we select three baselines for comparison. We choose ultrasound-based speech enhancement methodologies, \ie, Ultraspeech \cite{ding2022ultraspeech} and UltraSE \cite{sun2021ultrase}. We utilize the complex physical SE dataset (\ie, including the imaginary part of spectrograms) to evaluate all baselines because the inputs are complex spectrograms. Specifically, we adjust the input layer of Ultraspeech to adapt the input shape and only implement the enhancement component. In addition, we choose a speech-only baseline PHASEN \cite{yin2020phasen} to show the effectiveness of the involvement of ultrasound. The overview of the model design is as follows:

\begin{itemize}
    \item \textbf{Ultraspeech} \cite{ding2022ultraspeech}: The model utilizes a two-branch neural network for speech enhancement, combining ultrasound and speech signals. It uses a complex ratio mask to estimate noisy speech's magnitude and phase components. The interaction module allows information exchange between ultrasound and speech branches, improving noise discrimination.

    \item \textbf{UltraSE} \cite{sun2021ultrase}: The model borrows a multi-modal design for single-channel speech enhancement using ultrasound and speech signals. It employs a two-stream architecture to process speech and ultrasound separately, followed by a self-attention fusion mechanism to integrate the features. The model also uses a cGAN-based cross-modal training model to enhance the noisy speech spectrograms.

    \item \textbf{PHASEN} \cite{yin2020phasen}: The model features two parallel streams for amplitude and phase prediction. The amplitude stream includes frequency transformation blocks to capture global frequency correlations, while the phase stream benefits from information exchange between the streams, allowing the model to enhance both amplitude and phase.
    
\end{itemize}

\begin{figure}[t]
    \centering
    \includegraphics[width=1\linewidth]{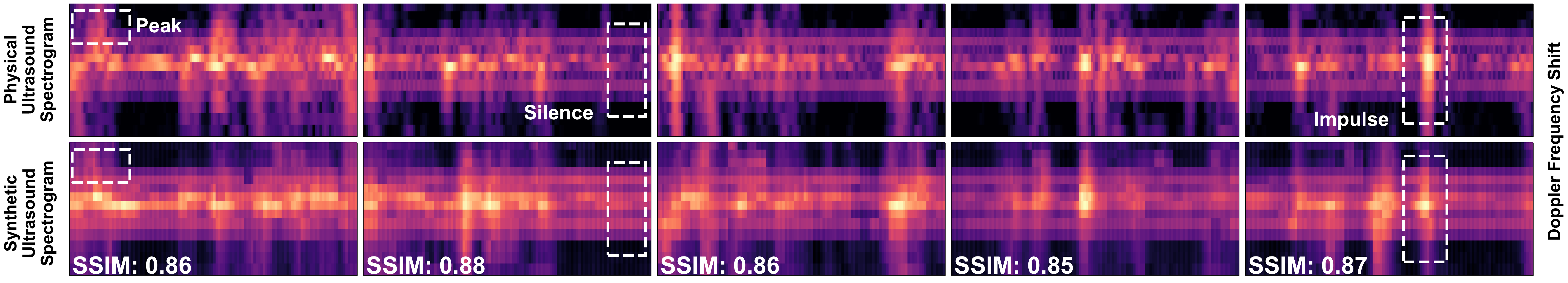}
    \caption{Visualization of the physical \vs \sysname synthetic ultrasound spectrograms. \sysname is capable of synthesizing high-quality ultrasound spectrograms, particularly when integrating video information. We calculate the single-spectrogram Structural Similarity Index Measure (\textbf{SSIM}) between the physical and \sysname synthetic ultrasound spectrograms per sample, providing a quantitative measure of synthesis fidelity.}
    \label{fig:visualization}
\end{figure}

\subsection{Evaluation Metrics}
We use three quantitative metrics (PESQ, STOI, and LSD) and one qualitative metric (MOS) to evaluate the quality of enhanced speech.

\noindent\textbf{PESQ}: \underline{P}erceptual \underline{E}valuation of \underline{S}peech \underline{Q}uality \cite{itu_p862_2} is an objective standard for evaluating the quality of speech standardized by ITU-T P.862. PESQ returns a score from -0.5 to 4.5, with higher scores indicating better quality.

\noindent\textbf{STOI}: \underline{S}hort-\underline{T}ime \underline{O}bjective \underline{I}ntelligibility \cite{taal2010short} is an objective metric used to assess the intelligibility of the speech signals which predicts how understandable speech is in various conditions, particularly when the speech is affected by noise and distortion. STOI returns a score from 0 to 1, with higher scores meaning better quality.

\noindent\textbf{LSD}: \underline{L}og-\underline{S}pectal \underline{D}istance \cite{rabiner1999fundamentals} is used to quantify the difference or distortion between two audio signals in the frequency domain, shown on decibels scale. LSD score is better if lower.

\noindent\textbf{MOS}: \underline{M}ean \underline{O}pinion \underline{S}core \cite{rothauser1969ieee} is a subjective measurement that is used to evaluate the quality of human-computer interactions. In a typical MOS test, a group of human subjects is asked to rate the quality of media samples on a scale, usually ranging from 1 to 5 (the higher, the better quality), shown in the Appendix \tab \ref{tab:mos_opinion_score}. We recruit 10 volunteers to conduct a qualitative evaluation of the enhanced speech using MOS. In experiments, we first play the clean speech, then the noisy, and its corresponding enhanced speech without knowing whether the enhanced speech from \sysname. Each volunteer will be assigned 100 samples (20 for each method) randomly chosen from different SNRs. The final results will be averaged for all volunteers.

\section{Experimental Evaluation}
\label{sec:experimental_evaluation}
\label{sec:exp}

\subsection{Overall Performance}
\label{sec:overall_performance}

\subsubsection{Visualization Analysis}
\label{sec:visualization_analysis}
As shown in Figure \ref{fig:visualization}, the synthetic ultrasound spectrogram visually exhibits a striking similarity to the corresponding physical ultrasound spectrogram in the first column. Additionally, we provide a quantitative evaluation of the synthesis fidelity across samples. We use the Structural Similarity Index Measure (SSIM) \cite{wang2004image}, a widely recognized metric that assesses similarity by analyzing the means and covariances of two samples, to evaluate the similarity between the synthetic and real ultrasound spectrograms. Each column shows the corresponding samples, with the SSIM value of the synthetic ultrasound spectrogram displayed at the bottom left. For example, the first spectrogram shows a fidelity of 0.86 compared with the physical spectrogram. We can clearly see that the first peak is generated at the correct temporal interval, well aligned with the physical one. In the second sample, the synthesis model accurately generates the silence at the correct time interval. In the last sample, we demonstrate that the model can also generate the impulse of the ultrasound spectrogram. In a nutshell, the synthesis model is capable of generating high-quality ultrasound spectrograms even in different patterns, \eg, peak, silence, impulse, \etc.

\begin{table}[t]
    \caption{\rev{Quantitative and qualitative evaluation results. The $\Delta$ metrics show the improvement or reduction compared with the Noisy SE Dataset. \sysname w/ phy. or syn. means the speech enhancement network of \sysname is trained on the synthetic or physical SE dataset and evaluated on the physical SE dataset.}}
    \centering
    \begin{tabular}{c|cccccc|c}
    \toprule
            \textbf{Method} & \textbf{PESQ $\uparrow$} & $\Delta$PESQ $\uparrow$ & \textbf{STOI $\uparrow$} & $\Delta$STOI $\uparrow$ & \textbf{LSD $\downarrow$} & $\Delta$ LSD $\downarrow$ & \textbf{MOS $\uparrow$} \\
         \cmidrule{1-8}
         Noisy SE Dataset & 2.33 & - & 0.77 & - & 1.16 & - & - \\
         \cmidrule{1-8}
         \textbf{\sysname} w/ phy. & 3.19 & +36.9\% & 0.89 & +15.6\% & 0.75 & -35.3\% & 4.52 \\
         \textbf{\sysname} w/ syn. & 3.10 & +33.0\% & 0.88 & +14.3\% & 0.80 & -31.0\% & 4.48 \\
         Ultraspeech \cite{ding2022ultraspeech} & 2.07 & -11.2\% & 0.73 & -5.2\% & 1.49 & +28.4\% & 2.24 \\
         UltraSE \cite{sun2021ultrase} & 2.23 & -4.3\% & 0.76 & -1.3\% & 1.33 & +14.7\% & 2.76 \\
         PHASEN \cite{yin2020phasen} & 3.02 & +29.6\% & 0.86 & +11.7\% & 1.04 & -10.3\% & 3.98 \\
    \bottomrule
    \end{tabular}
    \label{tab:quantitative_eva}
\end{table}

\subsubsection{Quantitative Evaluation}
\label{sec:quantitative_evaluation}

We first demonstrate the capability of the speech enhancement network by comparing it with other baselines, and based on this fact, we further validate the effectiveness of the ultrasound synthesis framework. 

As illustrated in \tab \ref{tab:quantitative_eva}, we elaborate on the results of \sysname compared with others. Compared with other baselines trained on the physical SE dataset, \sysname w/ phy. outperforms in all four metrics, with a PESQ gain of 36.9\%, an STOI gain of 15.6\%, and an LSD drop from 1.16 for the noisy dataset to 0.75. The results indicate that the proposed enhancement network is capable of improving both quality and intelligibility. While both ultrasound-based speech enhancement Ultraspeech \cite{ding2022ultraspeech} and UltraSE \cite{sun2021ultrase} fail to enhance our noisy SE dataset with human-related nonspeech noise. For Ultraspeech, they only utilize another noise dataset \cite{hu2010tandem} that only contains ambient noise \eg, musical instruments and machines (\ie, no human nonspeech). We observe that Ultraspeech also amplifies human-related noise, indicating it might not perform as expected when enhancing the human-related noise sources \cite{rashid2023nonspeech7k} we utilized. For UltraSE, they design a complex model with parameters $\sim 286$ M, which is too complex for speech enhancement. They do not release their code or adversarial hyperparameters, so we implemented the neural network and set the proper margin of Triplet loss; however, the model still fails to work effectively. The possible reason is that the model is too complex to enhance speech with the small-scale dataset—the physical SE dataset (2.4 hours)—compared with theirs (about 11 hours), especially the architecture of the discriminator. Furthermore, compared with their original result (STOI 0.80 and PESQ 3.01), \sysname can still achieve an STOI gain of 0.09 and maintain a PESQ of 3.19 when trained and evaluated on a much smaller dataset, indicating the effectiveness and efficiency of the proposed enhancement network. In addition, compared with PHASEN \cite{yin2020phasen}, \sysname w/ phy. achieves a PESQ score of 3.19, which is 0.17 higher, and a higher STOI score of 0.89, indicating the proposed enhancement network outperforms in intelligibility. Moreover, \sysname w/ phy. achieves a larger LSD reduction (-35.3\%) compared with only -10.3\% for PHASEN, showing that the enhancement network of \sysname performs remarkably well not only in the improvement of intelligibility but also in the quality of spectrograms. When comparing the enhancement network of \sysname trained with synthetic data (\sysname w/ syn.) to \sysname trained with physical data (\sysname w/ phy.), we find that the enhancement network trained on the synthetic SE dataset achieves comparable performance to that of the system trained on physical data. Specifically, \sysname w/ syn. achieves a PESQ score of 3.10, an STOI score of 0.88, and an LSD of 0.80. These results closely match those of \sysname w/ phy., which achieves a PESQ of 3.19, an STOI of 0.89, and an LSD of 0.75. The similarity in PESQ and STOI scores suggests that the synthetic ultrasound spectrograms are effective for speech enhancement tasks, while the close LSD values indicate high quality in the frequency domain of the synthetic spectrograms. 

Overall, these observations demonstrate that our ultrasound synthesis framework is capable of generating high-quality ultrasound spectrograms that are nearly as effective for speech enhancement as those derived from the physical dataset, confirming the robustness and practicality of using synthetic data.

\subsubsection{Qualitative Evaluation}
\label{sec:qualitative_evaluation}
In addition to the quantitative evaluation, we perform a qualitative evaluation using the MOS to assess the perceptual quality of the enhanced speech from a listener's perspective. As shown in \tab \ref{tab:quantitative_eva}, \sysname w/ phy. achieves a MOS of 4.52, significantly outperforming the baselines such as Ultraspeech (2.24), UltraSE (2.76), and PHASEN (3.98). The results indicate that listeners perceive the speech enhanced by \sysname w/ phy. as much clearer and more intelligible compared to other methods. The superiority of \sysname w/ phy. in terms of MOS, along with the quantitative gains in PESQ, STOI, and LSD, demonstrates that the proposed enhancement network delivers a noticeably higher perceptual quality of recovered speech than the other baselines. Comparing \sysname w/ syn. and phy., we observe that \sysname w/ syn. achieves a comparable MOS of 4.48, closely matching the perceptual quality of \sysname trained on the physical SE dataset. This close result between the synthetic and physical SE datasets supports the effectiveness of the synthesis framework. It shows that the synthesis model that can generate the ultrasound spectrograms is nearly indistinguishable from the model generated using physical data, proving the robustness and practicality of our synthetic data generation approach.

\begin{table}[h]
    \centering
    \caption{\rev{Quantitative evaluation results after using ultrasonic large-scale datasets as the training set.}}
    \label{tab:largescale_dataset}
    \begin{tabular}{c|c|cccccc}
    \toprule
         \textbf{Dataset} & \textbf{Scale} & \textbf{PESQ $\uparrow$} & $\Delta$PESQ $\uparrow$ & \textbf{STOI $\uparrow$} & $\Delta$STOI $\uparrow$ & \textbf{LSD $\downarrow$} & $\Delta$LSD $\downarrow$ \\
    \cmidrule{1-8}
        \rowcolor{red!20} {Noisy LJSpeech} & - & 1.83 & - & 0.78 & - & 1.67 & - \\
        \cmidrule{1-8}
         \multirow{4}{*}{\sysname w/ LJSpeech} 
         & $\quartercirc_{6h}$ & 3.03 & +65.6\% & 0.86 & +10.3\% & 1.02 & -38.9\% \\
         & $\halfcirc_{12h}$ & 3.10 & +69.4\% & 0.87 & +11.5\% & 0.94 & -43.7\% \\
         & $\threequartercirc_{18h}$ & 3.19 & +74.3\% & 0.87 & +11.5\% & 0.92 & -44.9\% \\
         & $\fullcirc_{24h}$ & 3.21 & +75.4\% & 0.88 & +12.8\% & 0.91 & -45.5\% \\
        \cmidrule{1-8}
        
        \rowcolor{NavyBlue!30} {Noisy TIMIT} & - & 2.01 & - & 0.77 & - & 1.32 & - \\
        \cmidrule{1-8}
        \multirow{4}{*}{\sysname w/ TIMIT} 
        & $\quartercirc_{1.45h}$ & 2.95 & +46.8\% & 0.81 & +5.2\% & 1.22 & -7.6\% \\
        & $\halfcirc_{2.90h}$ & 2.98 & +48.3\% & 0.85 & +10.4\% & 0.94 & -28.8\% \\
        & $\threequartercirc_{4.35h}$ & 3.08 & +53.2\% & 0.86 & +11.7\% & 0.86 & -34.8\% \\
        & $\fullcirc_{5.80h}$ & 3.11 & +55.0\% & 0.86 & +11.7\% & 0.85 & -36.4\% \\
        \cmidrule{1-8}
        
        \rowcolor{SeaGreen!40} {Noisy VCTK} & - & 1.90 & - & 0.69 & - & 2.03 & - \\
        \cmidrule{1-8}
        \multirow{4}{*}{\sysname w/ VCTK} 
        & $\quartercirc_{11h}$ & 3.17 & +66.8\% & 0.89 & +29.0\% & 0.79 & -61.1\% \\
        & $\halfcirc_{22h}$ & 3.26 & +71.6\% & 0.90 & +30.4\% & 0.77 & -62.1\% \\
        & $\threequartercirc_{33h}$ & 3.28 & +72.6\% & 0.92 & +33.3\% & 0.75 & -63.0\% \\
        & $\fullcirc_{44h}$ & 3.30 & +73.7\% & 0.92 & +33.3\% & 0.75 & -63.0\% \\
    \bottomrule
    \end{tabular}
\end{table}

\subsection{Large-scale Ultrasonic Datasets}
\label{sec:large-scale_dataset}
\rev{
To further evaluate the effectiveness of the synthesis framework and the generalizability of the enhancement network, we employ three large-scale synthetic SE datasets constructed from distinct, publicly available speech-only datasets: LJSpeech \cite{ljspeech17}, TIMIT \cite{garofolo1993timit}, and VCTK \cite{yamagishi2012}. The synthesis model, pre-trained on the ultrasound synthesis dataset, generates ultrasonic large-scale datasets. The large-scale synthetic datasets are exclusively used as training sets to ensure effectiveness, while the test set from the physical SE dataset is reserved for evaluation. The dataset size is incrementally increased from a $25\%$ subset to the full dataset. 
\tab \ref{tab:largescale_dataset} demonstrates that \sysname consistently achieves significant improvements across multiple metrics (PESQ, STOI, and LSD) as the training dataset size grows. For example, with only $25\%$ of the VCTK dataset, \sysname attains comparable performance to the 2.4-hour physical SE dataset, showing that even relatively small portions of the synthetic ultrasonic large-scale datasets yield substantial improvements in speech quality and intelligibility. Moreover, scaling the VCTK dataset from $25\%$ to the full dataset boosts PESQ from 1.90 to 3.30 and reduces LSD by more than half, underscoring how increased speaker diversity and dataset volume enhance speech intelligibility (STOI) and quality (PESQ). Although the LJSpeech corpus features fewer speakers, it exhibits an impressive $69.4\%$ PESQ improvement when scaled up to the 12-hour training set. Conversely, the TIMIT dataset, which includes many speakers with short utterances, achieves a $53.2\%$ PESQ improvement at full scale, reinforcing the synthesis model's ability to generalize across datasets with varying speaker counts and utterance lengths.

In summary, these results validate the generalizability and effectiveness of the proposed synthetic framework. As the training corpus size increases, performance steadily improves, and even small subsets of large corpora outperform the original manually collected physical SE dataset. The framework proves to be robust across diverse speech datasets, highlighting its utility for large-scale speech enhancement tasks.
}

\subsection{Different Noise Interference}
\label{sec:different_noise_interference}

\begin{table}[!htbp]
    \caption{\rev{Quantitative performance of \sysname in different noise interference, including environmental interference, human voice interference, mixed interference, and competing speakers interference. The noisy datasets are shown in different colors.}}
    \centering
    \label{tab:other_dataset_interference}
    \begin{tabular}{c|cccccc}
    \toprule
         \textbf{Method} & \textbf{PESQ $\uparrow$} & $\Delta$PESQ $\uparrow$ & \textbf{STOI $\uparrow$} & $\Delta$STOI $\uparrow$ & \textbf{LSD $\downarrow$} & $\Delta$ LSD $\downarrow$ \\
         \cmidrule{1-7}

         \rowcolor{gray!20} \textbf{Environmental} & 2.10 & - & 0.74 & - & 1.42 & - \\
         \cmidrule{1-7}
         \sysname w/ phy. & 3.31 & +57.6\% & 0.90 & +21.6\% & 0.70 & -50.7\% \\
         \sysname w/ syn. & 3.18 & +51.4\% & 0.89 & +20.3\% & 0.73 & -48.6\% \\
         Ultraspeech & 2.83 & +34.8\% & 0.78 & +5.4\% & 0.98 & -31.0\% \\
         UltraSE & 2.81 & +33.8\% & 0.76 & +2.7\% & 1.26 & -11.3\% \\

         \cmidrule{1-7}
         \rowcolor{NavyBlue!30} \textbf{Human Voice} & 2.27 & - & 0.74 & - & 1.48 & - \\
         \cmidrule{1-7}
         \sysname w/ phy. & 3.22 & +41.9\% & 0.87 & +17.6\% & 0.75 & -49.3\% \\
         \sysname w/ syn. & 3.18 & +40.1\% & 0.86 & +16.2\% & 0.80 & -45.9\% \\
         Ultraspeech & 2.17 & -4.4\% & 0.74 & +0.0\% & 1.52 & +2.7\% \\
         UltraSE & 2.64 & +16.3\% & 0.79 & +6.8\% & 1.15 & -22.3\% \\

         \cmidrule{1-7}
         \rowcolor{SeaGreen!40} \textbf{Mixed} & 2.32 & - & 0.75 & - & 1.38 & - \\
         \cmidrule{1-7}
         \sysname w/ phy. & 2.95 & +27.2\% & 0.85 & +13.3\% & 0.87 & -36.9\% \\
         \sysname w/ syn. & 2.88 & +24.1\% & 0.83 & +10.7\% & 0.92 & -33.3\% \\
         Ultraspeech & 2.33 & +0.4\% & 0.73 & -2.7\% & 1.41 & +2.2\% \\
         UltraSE & 2.54 & +9.5\% & 0.79 & +5.3\% & 1.16 & -15.9\% \\

         \cmidrule{1-7}
         \rowcolor{red!20} \textbf{Competing Speakers} & 2.29 & - & 0.75 & - & 1.43 & - \\
         \cmidrule{1-7}
         \sysname w/ phy. & 2.82 & +23.1\% & 0.83 & +10.7\% & 1.03 & -27.9\% \\
         \sysname w/ syn. & 2.78 & +21.4\% & 0.81 & +8.0\% & 1.06 & -25.9\% \\
         Ultraspeech & 2.09 & -8.7\% & 0.72 & -4.0\% & 1.48 & +3.5\% \\
         UltraSE & 2.64 & +15.3\% & 0.78 & +4.0\% & 1.29 & -9.8\% \\

    \bottomrule
    \end{tabular}
\end{table}

\rev{
We compare our speech enhancement network (\sysname) and other baselines \cite{sun2021ultrase, ding2022ultraspeech} across four different interference types: environmental, human voice, mixed, and competing speakers, to evaluate their robustness and adaptability to diverse noise distributions. The noisy datasets are handcrafted as follows:
\begin{itemize}
    \item \textbf{Environmental Interference}: The ESC-50 \cite{piczak2015esc} dataset, containing various ambient noises, is mixed with the collected dataset at SNRs of \{-10, -5, 0, 5, 10, 15\} dB.
    \item \textbf{Human Voice Interference}: The TIMIT \cite{garofolo1993timit} corpus is blended with the collected dataset, using SNRs of \{-10, -5, 0, 5, 10\} dB.
    \item \textbf{Mixed Interference}: A combination of AudioSet \cite{gemmeke2017audio}, ESC-50 \cite{piczak2015esc}, and TIMIT is mixed at SNRs of \{-10, -5, 0, 5, 10, 15\} dB.
    \item \textbf{Competing Speakers Interference}: Speech samples from different users are overlayed with the collected dataset at SNRs of \{0, 5, 10, 15\} dB.
\end{itemize}

The results in \tab \ref{tab:other_dataset_interference} show that \sysname demonstrates strong robustness across all noise scenarios. 
Taking environmental interference as an example, \sysname w/ syn. achieves a 51.4\% gain in PESQ, a 20.3\% improvement in STOI, and a 48.6\% reduction in LSD relative to the noisy baseline. \sysname w/ phy. achieves even better results with a 57.6\% PESQ gain, 21.6\% STOI improvement, and 50.7\% LSD reduction. Ultraspeech and UltraSE show moderate improvements but consistently fall short of \sysname.
Similar performance gains are obtained for both human voice interference and mixed interference. 
Overall, \sysname with synthetic data achieves comparable performance to that with physical data, both significantly outperforming Ultraspeech and UltraSE.
}

\subsection{Dataset Scale Evaluation}
\label{sec:dataset_scale_evaluation}

\begin{table}[h]
    \centering
    \caption{Evaluation of the synthesis framework at different ultrasound synthesis dataset scales ($\fullcirc = 2.40h$).}
    \begin{tabular}{c|c|cccccc}
        \toprule
         \textbf{Dataset} & \textbf{Scale} & \textbf{PESQ $\uparrow$} & $\Delta$PESQ $\uparrow$ & \textbf{STOI $\uparrow$} & $\Delta$STOI $\uparrow$ & \textbf{LSD $\downarrow$} & $\Delta$LSD $\downarrow$ \\
         \cmidrule{1-8}
         Noisy SE Dataset & - & 2.33 & - & 0.77 & - & 1.16 & - \\
         \cmidrule{1-8}
         \multirow{4}{*}{\shortstack{Ultrasound Synthesis Dataset}}  
         & \quartercirc$_{0.6h}$ & \pesqtimebar{3.1}{1.28} & -45.1\% & \stoitimebar{0.88}{0.58} & -24.7\% & \lsdtimebar{1.57}{1.57} & +35.3\% \\
         
         & \halfcirc$_{1.2h}$ & \pesqtimebar{3.1}{2.03} & -12.9\% & \stoitimebar{0.88}{0.79} & +2.6\% & \lsdtimebar{1.57}{0.92} & -20.7\% \\
         
         & \threequartercirc$_{1.8h}$ & \pesqtimebar{3.1}{2.92} & +25.3\% & \stoitimebar{0.88}{0.86} & +11.7\% & \lsdtimebar{1.57}{0.86} & -25.9\% \\
         
         & \fullcirc$_{2.4h}$ & \pesqtimebar{3.1}{3.10} & +33.0\% & \stoitimebar{0.88}{0.88} & +14.3\% & \lsdtimebar{1.57}{0.80} & -31.0\% \\
         \bottomrule
    \end{tabular}
    
    \label{tab:scale_syn}
\end{table}

\begin{table}[h]
    \centering
    \caption{\rev{Evaluation of the enhancement network at different physical and synthetic SE dataset scales ($\fullcirc = 2.40h$).}}
    \label{tab:scale_en}
    \begin{tabular}{c|c|cccccc}
        \toprule
         \textbf{Dataset} & \textbf{Scale} & \textbf{PESQ $\uparrow$} & $\Delta$PESQ $\uparrow$ & \textbf{STOI $\uparrow$} & $\Delta$STOI $\uparrow$ & \textbf{LSD $\downarrow$} & $\Delta$LSD $\downarrow$ \\
         \cmidrule{1-8}
         Noisy SE Dataset & - & {2.33} & - & {0.77} & - & {1.16} & - \\
         \cmidrule{1-8}
         \multirow{4}{*}{\shortstack{Physical SE Dataset}} 
         & \quartercirc$_{0.6h}$ & \pesqtimebar{3.19}{2.90} & +24.5\% & \stoitimebar{0.89}{0.84} & +9.1\% & \lsdtimebar{0.85}{0.85} & -26.7\% \\
         & \halfcirc$_{1.2h}$ & \pesqtimebar{3.19}{2.95} & +26.6\% & \stoitimebar{0.89}{0.87} & +13.0\% & \lsdtimebar{0.85}{0.79} & -31.9\% \\
         & \threequartercirc$_{1.8h}$ & \pesqtimebar{3.19}{3.14} & +34.8\% & \stoitimebar{0.89}{0.88} & +14.3\% & \lsdtimebar{0.85}{0.73} & -37.1\% \\
         & \fullcirc$_{2.4h}$ & \pesqtimebar{3.19}{3.19} & +36.9\% & \stoitimebar{0.89}{0.89} & +15.6\% & \lsdtimebar{0.85}{0.75} & -35.3\% \\

         \cmidrule{1-8}
         \multirow{4}{*}{\shortstack{Synthetic SE Dataset}} 
         & \quartercirc$_{0.6h}$ & \pesqtimebar{3.1}{2.87} & +23.2\% & \stoitimebar{0.88}{0.83} & +7.8\% & \lsdtimebar{0.88}{0.88} & -24.1\% \\
         & \halfcirc$_{1.2h}$ & \pesqtimebar{3.1}{2.92} & +25.3\% & \stoitimebar{0.88}{0.85} & +10.4\% & \lsdtimebar{0.88}{0.85} & -26.7\% \\
         & \threequartercirc$_{1.8h}$ & \pesqtimebar{3.1}{3.08} & +32.2\% & \stoitimebar{0.88}{0.88} & +14.3\% & \lsdtimebar{0.88}{0.81} & -30.2\% \\
         & \fullcirc$_{2.4h}$ & \pesqtimebar{3.1}{3.10} & +33.0\% & \stoitimebar{0.88}{0.88} & +14.3\% & \lsdtimebar{0.88}{0.80} & -31.0\% \\
         
         \bottomrule
    \end{tabular}
\end{table}

\rev{
To better understand the relationship between performance and dataset scale in both the ultrasound synthesis framework and the speech enhancement network of \sysname, we conduct comprehensive evaluations. For the synthesis framework, we incrementally vary the size of the ultrasound synthesis dataset (2.4 hours) from 25\% to 100\% of the full dataset to train the synthesis model. For each trial, the trained synthesis model generates the training set of the synthetic SE datasets, which are utilized in the training phase of the enhancement network. Conversely, for the enhancement network, we keep the ultrasound synthesis dataset at full scale, vary the size of the synthetic and physical SE datasets (2.4 hours) in the same manner, and use these subsets to train and evaluate the enhancement network.

\tab \ref{tab:scale_syn} shows the performance of the synthesis framework with varying dataset scales. As expected, the metrics improve consistently as the dataset size increases. For example, PESQ improves from 1.28 at the 0.6-hour scale to 3.10 at the full 2.4-hour scale, and LSD decreases from 1.57 to 0.80, reflecting a significant enhancement in intelligibility and quality. Notably, the improvement becomes substantial when scaling beyond 1.2 hours, with STOI increasing from 0.79 to 0.88. These results emphasize the importance of leveraging larger-scale datasets for training the synthesis framework.

\tab \ref{tab:scale_en} illustrates the performance of the enhancement network trained on synthetic and physical datasets. When using synthetic SE datasets, the enhancement network achieves comparable results to the physical datasets, with PESQ scores of 3.10 and 3.19, and STOI scores of 0.88 and 0.89, respectively. At 75\% of the scale (1.8 hours), the synthetic dataset achieves performance similar to the full physical dataset, highlighting the effectiveness of the synthesis framework in generating high-quality ultrasound spectrograms. Furthermore, even with only the 0.6-hour subset of data, the synthetic dataset achieves impressive results, with a PESQ of 2.87 and an STOI of 0.83. 

}

\subsection{Real-World Evaluation}
\label{sec:real_world_evaluation}
\begin{figure}[t]
    \centering
    \includegraphics[width=0.8\linewidth]{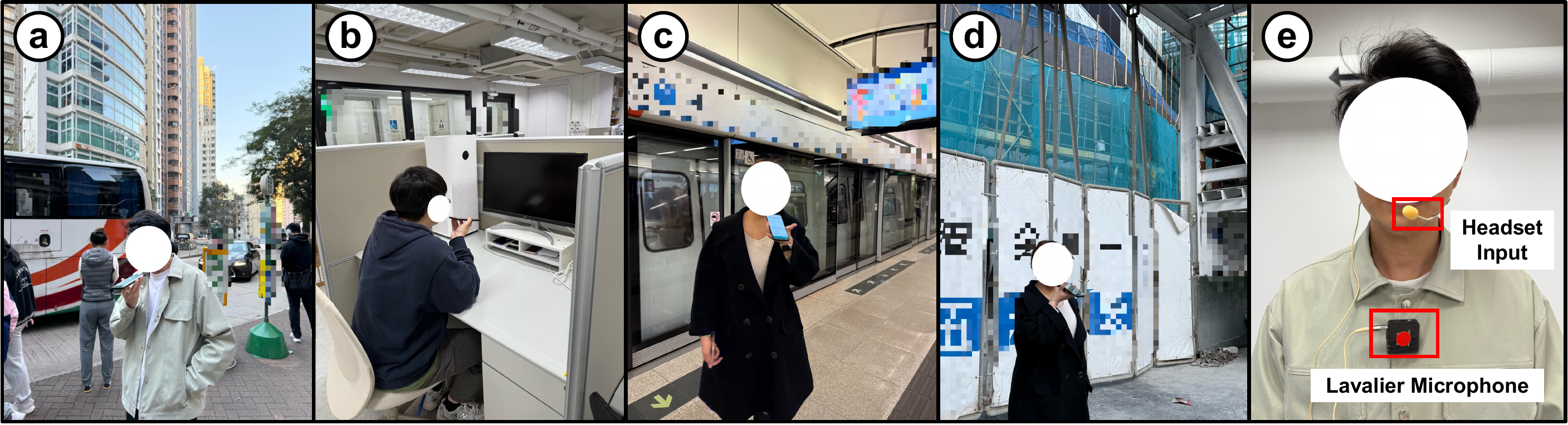}
    \caption{\rev{The illustration of real-world evaluation scenarios: (a) outdoor bus stop, (b) office room, (c) indoor metro station, and (d) construction site. (e) The ground truth recording setup uses a wireless lavalier microphone connected to a headset input.}}
    \label{fig:realworld_scenarios}
\end{figure}
\begin{table}[h]
    \centering
    \caption{\rev{The results of the end-to-end real-world evaluation with \sysname trained on various datasets, including the collected physical and synthetic SE datasets and the ultrasonic large-scale datasets (shown in different colors).}}
    \label{tab:real_world_results}
    \begin{tabular}{c|cccccc}
    \toprule
         \textbf{Training Dataset} & \textbf{PESQ $\uparrow$} & $\Delta$PESQ $\uparrow$ & \textbf{STOI $\uparrow$} & $\Delta$STOI $\uparrow$ & \textbf{LSD $\downarrow$} & $\Delta$LSD $\downarrow$\\
        \cmidrule{1-7}
        {Real-world Noisy Dataset} & 2.00 & - & 0.62 & - & 1.64 & -\\
        \cmidrule{1-7}
        {Physical SE Dataset} & 3.13 & +56.5\% & 0.85 & +37.1\% & 0.78 & -52.4\% \\
        {Synthetic SE Dataset} & 3.06 & +53.0\% & 0.83 & +33.9\% & 0.84 & -48.8\% \\
        \cmidrule{1-7}
        \rowcolor{red!20} {Ultrasonic LJSpeech} & 3.14 & +57.0\% & 0.86 & +38.7\% & 0.94 & -42.7\% \\
        \rowcolor{NavyBlue!20} {Ultrasonic TIMIT} & 3.06 & +53.0\% & 0.82 & +32.3\% & 0.91 & -44.5\% \\
        \rowcolor{SeaGreen!30} {Ultrasonic VCTK} & 3.29 & +64.5\% & 0.88 & +41.9\% & 0.78 & -52.4\% \\
    \bottomrule
    \end{tabular}
\end{table}

\rev{
We conduct an end-to-end real-world evaluation to assess \sysname in real-world settings. Three volunteers (2 males and 1 female) are recruited to read randomly selected materials for approximately 5 minutes. As shown in \fig \ref{fig:realworld_scenarios}e, we capture the ground truth data using a RODE Wireless GO II lavalier microphone \cite{rode}, connected to a headset input to ensure clean and accurate speech recording. The evaluation encompasses four distinct scenarios, as illustrated in \fig \ref{fig:realworld_scenarios}: 
1) \textbf{Outdoor bus stop}: Noise sources include human conversations, vehicle engines, and wind, with noise levels ranging from 80 to 85 dB.
2) \textbf{Office room}: Noise primarily originates from keyboard typing, door movements, and an air purifier, with levels between 50 and 60 dB.
3) \textbf{Indoor metro station}: Noise sources include train sounds, escalator engines, announcements, and human conversations, with levels ranging from 75 to 100 dB.
4) \textbf{Construction site}: Noise is primarily caused by construction activities and machine engine noise, with levels between 85 and 110 dB. 
The ambient noise levels were measured using a decibel meter.

The results of the end-to-end real-world evaluation are shown in \tab \ref{tab:real_world_results}. Both \sysname trained on the synthetic SE dataset and the physical SE dataset demonstrate substantial improvements over the noisy baseline across all metrics (PESQ, STOI, and LSD). The model trained on the synthetic SE dataset achieves a PESQ of 3.06 and an STOI of 0.83, reflecting a respective increase of 53.0\% and 33.9\% compared to the noisy dataset. Similarly, the physical SE dataset attains a PESQ of 3.13 and an STOI of 0.85, yielding a 56.5\% improvement in PESQ and a 37.1\% gain in STOI.
Additionally, we evaluate \sysname trained on the generated ultrasonic large-scale datasets (\eg, LJSpeech, TIMIT, and VCTK). All models trained on large-scale datasets demonstrate positive gains, confirming the validity of employing large-scale ultrasonic datasets for ultrasound-based speech enhancement. However, \sysname trained on LJSpeech shows timbre bias toward a female voice since the dataset contains only a single female speaker and lacks timbre diversity. 
TIMIT and VCTK datasets include diverse speakers, with 630 and 110 speakers, respectively. The limited utterance duration per speaker (approximately 30 seconds per speaker) in TIMIT results in a slight performance drop. On the other hand, VCTK balances speaker diversity and utterance duration per speaker, leading to the best performance among the large-scale datasets. 

\mrev{We note that our current evaluation scale is relatively limited due to the cost of synchronized data collection. 
Nevertheless, the overall results validate the practicality and scalability of the proposed system in real-world environments.} \sysname consistently delivers robust performance across diverse noise scenarios, making it a viable solution for challenging real-world speech enhancement tasks.
}

\begin{figure}[t]
    \subfloat[Physical SE dataset and synthesis SE dataset.]{
          \includegraphics[width=0.5\linewidth]{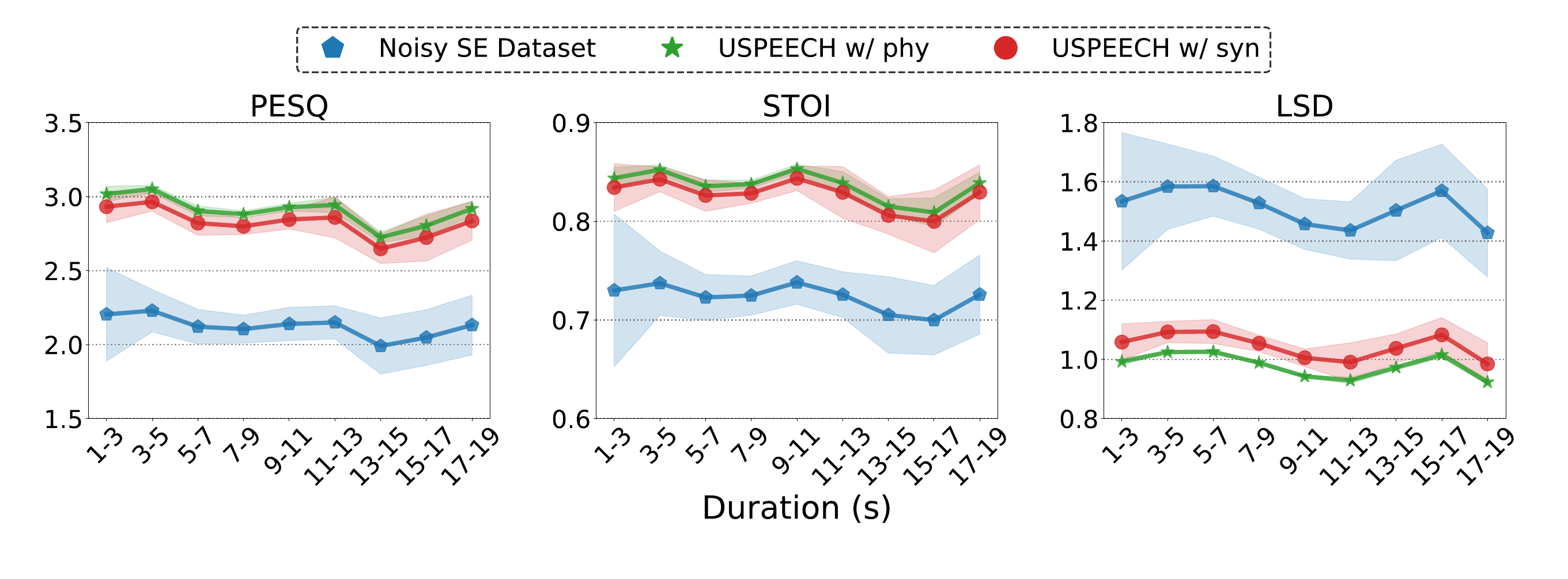}}
    \subfloat[Large-scale synthetic datasets.]{
      \includegraphics[width=0.5\linewidth]{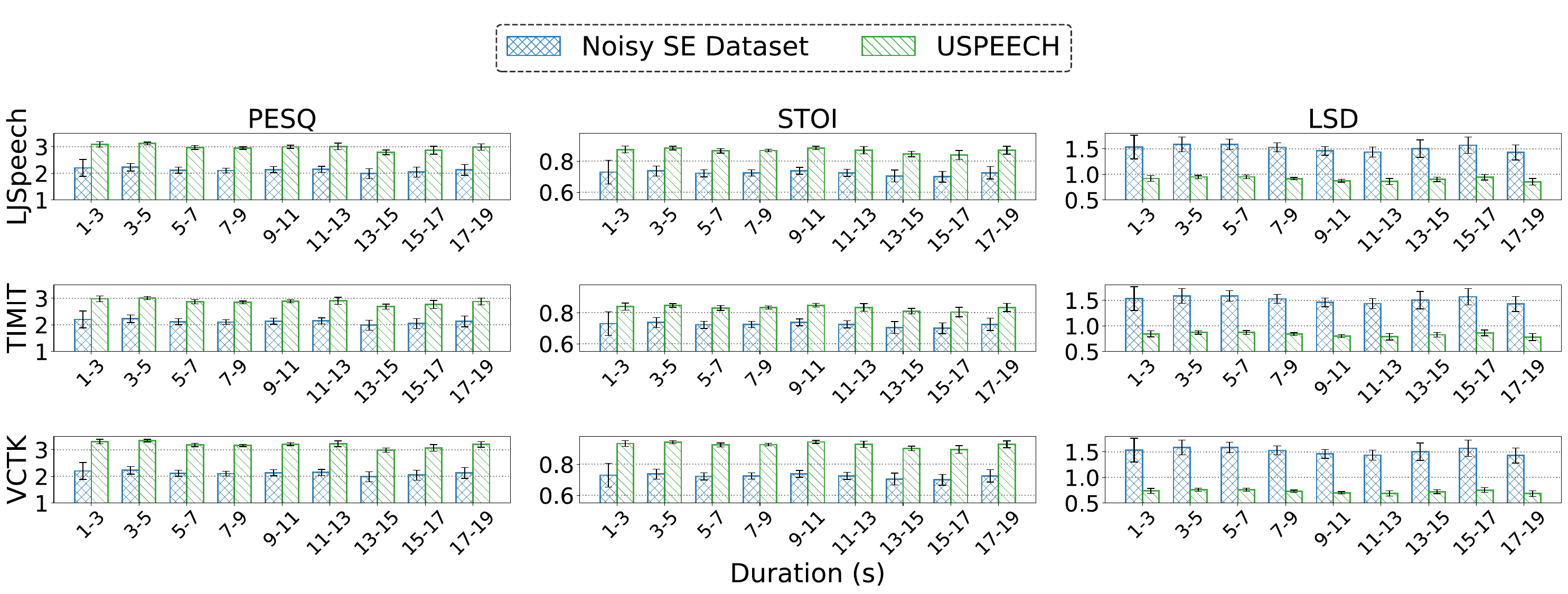}
   }
   \vspace{-0.1in}
    \caption{Performance on temporal variability.}
    \vspace{-0.1in}
 \label{fig:temporal}
\end{figure}

\begin{figure}[t]
    \centering\includegraphics[width=1\linewidth]{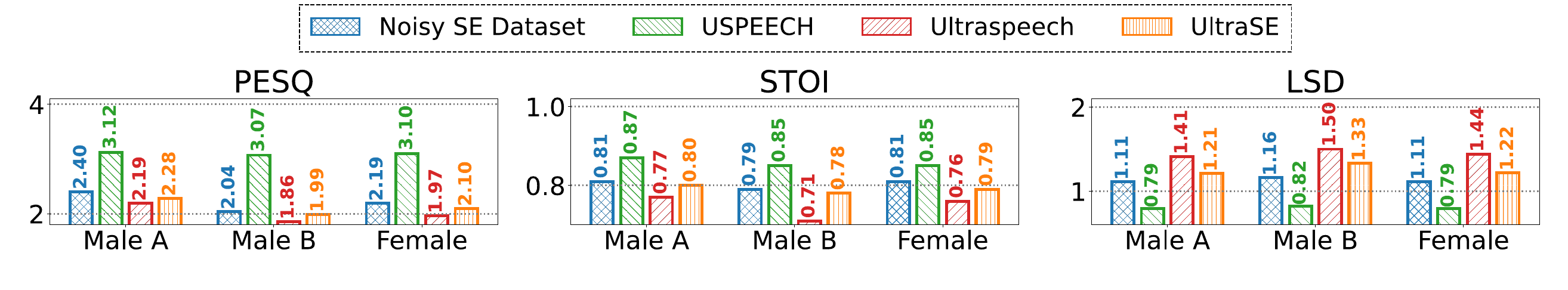}
    \vspace{-0.2in}
    \caption{Performance on unseen users.}
    \label{fig:unseen_user}
    \vspace{-0.2in}
\end{figure}

\subsection{Temporal Variability Performance}
\label{sec:temporal_variability}
\rev{
To demonstrate the adaptability and robustness of \sysname in handling variable-length input, we conduct experiments across multiple time intervals to evaluate the system's resilience to temporal variations, as shown in \fig \ref{fig:temporal}. These experiments are evaluated on the test set of the physical SE dataset with the model trained on the physical and synthetic SE dataset, as well as the ultrasonic large-scale synthetic SE datasets. In contrast to prior experiments that used fixed-length audio segments, this experiment introduces variability in input duration, simulating real-world conditions where speech length fluctuates dynamically.
The results across the physical SE dataset, synthetic SE dataset, and the ultrasonic large-scale datasets consistently show that \sysname maintains strong and stable performance across varying input lengths in all three metrics. For the physical SE dataset, \sysname w/ phy. achieves consistent PESQ values ranging from about 2.9 to 3.2, indicating that \sysname handles longer audio inputs without significant degradation in perceptual quality. Similarly, STOI values remain robust, staying close to 0.88 across all durations, while LSD remains low, demonstrating that the system retains good spectral accuracy regardless of the input length. Moreover, \sysname trained on the synthetic SE dataset (\sysname w/ syn.) performs comparably to \sysname w/ phy., showing that synthetic ultrasound spectrograms also maintain stability across temporal variations.
For the model trained on the large-scale synthetic SE datasets and evaluated on the test set of the physical SE dataset, \sysname exhibits similarly strong performance. For example, in the LJSpeech dataset, PESQ scores for different durations range from 3.1 to 3.3, showing minimal fluctuation. STOI scores remain above 0.87 across all temporal intervals, and LSD consistently stays below 0.9, showing the system’s capability to manage both short and long audio durations without degrading quality or intelligibility. The TIMIT and VCTK datasets display similar trends, with PESQ and STOI remaining stable across varying time intervals, and LSD values showing only slight variations, all within acceptable ranges.
}

\subsection{Unseen User Performance}
\label{sec:appendix_unseen_user_performance}

\rev{
To evaluate the generalizability to unseen users, we test \sysname on a dataset collected from three new volunteers (Male A, Male B, and one female) using the same data collection procedure. The unseen user dataset is evaluated directly without additional training, as shown in \fig \ref{fig:unseen_user}. 
\sysname demonstrates strong generalization across unseen users, consistently outperforming the noisy dataset and other baselines in all metrics. On average, \sysname improves PESQ by over 0.9, STOI by approximately 0.07, and reduces LSD by more than 0.3 compared to the noisy dataset. Competing methods, such as Ultraspeech and UltraSE, achieve lower improvements, with PESQ values under 2.3 and LSD reductions that remain over 1.2 for most users.
The results on the unseen dataset (PESQ: 3.10, STOI: 0.86, LSD: 0.80) closely align with those from the seen user scenario, where \sysname achieves 3.19, 0.89, and 0.75, respectively. These findings highlight \sysname's robust speech enhancement capabilities, achieving higher perceptual quality, intelligibility, confirming its ability to generalize effectively without additional training.
}

\subsection{Ablation Studies}
\label{sec:ablation_study}
\begin{table}[h]
    \centering
    \caption{The results of ablation studies.}
    \label{tab:ablation}
    \begin{tabular}{c|cccccc}
        \toprule
         \textbf{Method} & \textbf{PESQ $\uparrow$} & $\Delta$PESQ $\uparrow$ & \textbf{STOI $\uparrow$} & $\Delta$STOI $\uparrow$ & \textbf{LSD $\downarrow$} & $\Delta$LSD $\downarrow$ \\
         \cmidrule{1-7}
         Noisy SE Dataset & 2.33 & - & 0.77 & - & 1.16 & - \\
         \cmidrule{1-7}
         \sysname w/ syn. & 3.10 & +33.0\% & 0.88 & +14.3\% & 0.80 & -31.0\% \\
         w/o video pre-training & 2.32 & -0.4\% & 0.74 & -3.9\% & 1.04 & -10.3\% \\
         w/o temporal-semantic loss & 2.84 & +21.9\% & 0.79 & +2.6\% & 0.99 & -14.7\% \\
         \cmidrule{1-7}
         \sysname w/ phy. & 3.19 & +36.9\% & 0.89 & +15.6\% & 0.75 & -35.3\% \\
         w/o ultrasound branch & 2.43 & +4.3\% & 0.77 & 0.0\% & 0.79 & -31.9\% \\
         w/o neural vocoder & 2.90 & +24.5\% & 0.81 & +5.2\% & 0.94 & -19.0\%\\
         w/o dual-MSE loss & 3.15 & +35.2\% & 0.86 &+11.7\% & 0.78 &-32.8\% \\
         w/o FTLs & 3.01 & +29.2\% & 0.84 & +9.1\% & 0.89 & -23.3\% \\
         \bottomrule
    \end{tabular}
\end{table}
\rev{
We perform ablation studies to validate the effectiveness of the proposed components in \sysname. The corresponding results are shown in \tab \ref{tab:ablation}.

\textbf{w/o video pre-training} refers to skipping the contrastive video-audio pre-training phase described in \S \ref{sec:contrastive_pretraining} and directly training the synthesis model without incorporating the physical correspondence between video and ultrasound articulatory gestures. Without this pre-training, the synthesis framework experiences a significant drop in performance. These results highlight the necessity of video pre-training and the critical role of physical correspondence between video and ultrasound in generating high-quality synthetic ultrasound data.

\textbf{w/o temporal-semantic loss} refers to replacing the proposed temporal-semantic loss with the InfoNCE loss on video and audio embeddings. This leads to a performance reduction, with PESQ decreasing by 0.26, STOI by 0.09, and LSD increasing by 0.19, showing the importance of temporal-semantic alignment in preserving articulatory information.

\textbf{w/o ultrasound branch} refers to removing the ultrasound feature concatenation before the Transformer by replacing it with replicated speech features in the enhancement network, as described in \S \ref{sec:t_f_mel_spectrogram_enhancement}. The system trained on the physical SE dataset suffers a notable performance decline. These results clearly demonstrate the importance of integrating the ultrasound branch, which provides critical articulatory information to enhance speech intelligibility and quality.

\textbf{w/o neural vocoder} refers to replacing the vocoder with the GriffinLim algorithm to recover the waveform from the amplitude of spectrograms. This results in a performance decline, with PESQ dropping by 0.29 and LSD increasing by 0.19, showing the neural vocoder’s effectiveness in recovering high-quality waveforms.

\textbf{w/o dual-MSE loss} means removing the MSE loss of time differential spectrograms when training, leaving the MSE loss of the spectrogram (\ie, set $\alpha$ to 1). This reduces PESQ by 0.04 and increases LSD by 0.14, demonstrating the value of capturing both static and dynamic spectrogram features.

\textbf{w/o FTLs} indicates removing the Frequency Transformation Layers (FTLs) in skip connections in the enhancement network. This causes a 0.18 reduction in PESQ and a 0.09 reduction in STOI, emphasizing the role of FTLs in preserving harmonic correlations and improving spectrogram quality.
}

\subsection{Impact of Different Factors}
\label{sec:impact_of_different_factors}
\begin{figure}[h]
    \centering
    \includegraphics[width=\linewidth]{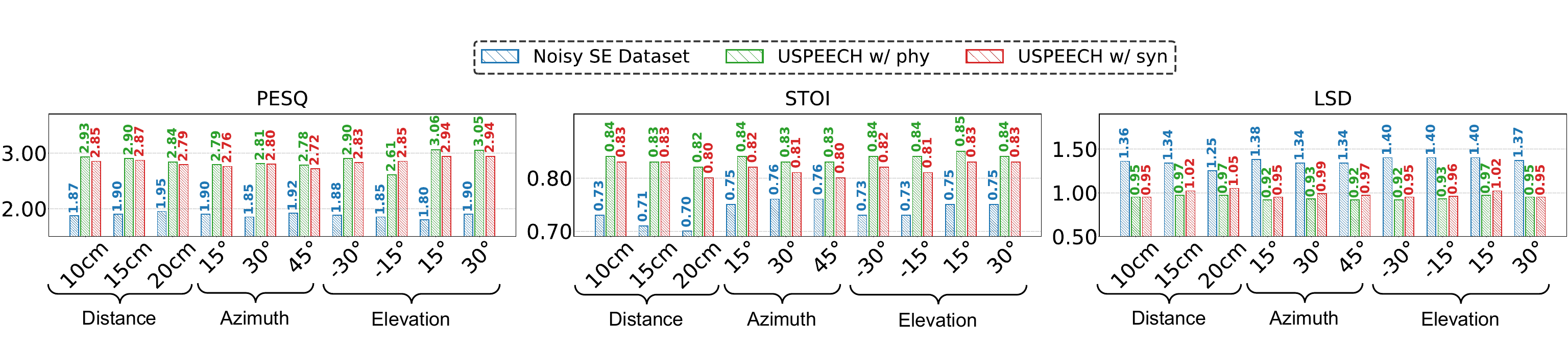}
    \caption{The performance under different distances, azimuths, and elevations.}
    \label{fig:factors}
    \vspace{-0.1in}
\end{figure}

To evaluate the robustness of \sysname under varying conditions, we conduct controlled experiments isolating three key factors: device distance, azimuth, and elevation, while keeping all other variables constant, as described in \S \ref{sec:data_collection}. A single volunteer reads a standardized passage\footnote{{https://www.bbc.com/news/technology-67379533}}, and all data collected are used exclusively for testing. The performance across these factors is shown in \fig \ref{fig:factors}.

\begin{itemize}
    \item \textbf{Distance}:  
    We test microphone distances of 10, 15, and 20 cm from the mouth. As the distance increases, there is a marginal decline in performance. For example, at 20 cm, \sysname w/ phy achieves a PESQ of 2.88, an STOI of 0.82, and an LSD of 1.05, compared to 2.94, 0.84, and 0.95 at 10 cm. Despite this, \sysname demonstrates robust performance across all distances.

    \item \textbf{Azimuth}: 
    The microphone is rotated to 15, 30, and 45 degrees relative to the mouth. \sysname maintains effective speech enhancement, with only slight variations in metrics. For instance, at 45 degrees, \sysname w/ phy achieves a PESQ of 2.77, an STOI of 0.80, and an LSD of 0.96, compared to 2.85, 0.82, and 0.95 at 15 degrees.

    \item \textbf{Elevation}:  
    Elevation angles of $\pm 15$ and $\pm 30$ degrees are tested, where negative angles face the mouth, and positive angles face the throat. Results show minimal performance differences, with \sysname achieving a PESQ of 2.93, an STOI of 0.83, and an LSD of 1.34 at +30 degrees, compared to 2.90, 0.84, and 1.32 at -30 degrees, demonstrating strong adaptability to elevation changes.
\end{itemize}

Overall, these experiments confirm that \sysname is robust to variations in device placement and orientation, with only minor performance reductions across all tested conditions.

\section{Discussion and Future Work}
\label{sec:discussion_and_future_work}

\head{Human Effort}
Collecting clean audio-ultrasound datasets is challenging due to environmental noise and ethical concerns. As mentioned above, audio can be disrupted by ambient sounds like footsteps or door movements, while ultrasound is sensitive to minor human actions like swallowing or head movements. The interference can overwhelm the desired signals, making it hard to capture clean data without a controlled environment, which requires significant effort and resources. Privacy concerns also add complexity, as many speech datasets cannot be publicly shared due to legal and ethical restrictions on personal data. To address these issues, \sysname minimizes the need for large-scale manual data collection by first gathering a small dataset to train a synthesis model. This model then synthesizes the larger clean ultrasound-audio datasets from publicly available sources, including LJSpeech \cite{ljspeech17}, TIMIT \cite{garofolo1993timit}, and VCTK \cite{yamagishi2012}. The \sysname reduces both human effort and interference while preventing privacy and ethical concerns tied to personal data collection.

\head{Potential Applications}
\sysname's ability to synthesize reliable large-scale audio-ultrasound datasets opens up opportunities for ultrasound-based sensing applications. One promising area is silent speech interfaces, where ultrasound captures articulatory movements without audible speech, benefiting individuals with speech impairments or enabling silent communication in noise-sensitive environments. Another application is hardware-free anti-eavesdropping mechanisms. By using ultrasound frequencies, speech information can be transmitted inaudibly, reducing the risk of unauthorized listening and enhancing privacy in sensitive conversations. Gesture recognition is also a potential application, where ultrasound signals detect hand gestures or body movements, enabling natural human-computer interaction. This could be applied in virtual reality, gaming, smart home control, and assistive technologies. By facilitating the generation of large-scale datasets, \sysname supports the training of machine learning models necessary for these applications, accelerating innovation and leading to more robust sensing systems. Future research could develop prototypes and conduct user studies to evaluate effectiveness and usability in these areas.

\head{Real-World Multi-User Scenarios}
In our experimental evaluation (\S \ref{sec:different_noise_interference}), we simulated multi-user scenarios such as competing speaker interference and human voice interference. However, we did not collect datasets from actual multi-user environments. Despite this, \sysname still enhances noisy speech signals effectively. For future work, collecting multi-user datasets and incorporating them into the training stage could lead to better performance in real-world applications. This involves challenges like overlapping speech and dynamic acoustic environments. Incorporating real-world data could enhance the system's ability to isolate and enhance target speech in complex settings. While the paper focuses on synthesizing large-scale audio-ultrasound datasets from small-scale datasets, addressing real-world multi-user speech separation remains an open issue for future research. Future work could explore advanced signal separation algorithms tailored to multi-user environments, such as multi-microphone array processing and deep learning-based speech separation techniques.

\section{Related Works}
\label{sec:related_works}

\subsection{Cross-modal Data Synthesis}
\label{sec:cross_modal_data_synthesis}
As the size of neural networks increases, so does the need for training data. Data synthesis is one way to alleviate the dilemma. Prior works proposed to synthesize data for human pose \cite{huamnposegeneration}, IMU \cite{imugeneration}, and depth images \cite{depthgeneration} from videos and motion capture (MoCap), which are readily available as open-source data. This idea also applies to wireless data synthesis, which generates RF sensing data from modalities like kinematic modeling \cite{kinematicmodaling1, kinematicmodaling2}, MoCap datasets \cite{mocap1, mocap3}, Kinect \cite{kinectradar1, kinectradar2}, IMU or video datasets \cite{xgait,Vid2Doppler}. \rev{However, to the best of our knowledge, there exists no prior work on ultrasound data synthesis.} 

\rev{
For speech-related cross-modal data synthesis, numerous studies explore video-audio generation due to their natural pairing relationship, where video-to-speech focuses on generating the missing speech from soundless videos by capturing lip motions \cite{LipVoicer}. Recent works \cite{Vid2speech,Lipper} investigate the potential of modeling talking faces as intelligible speech. To explore more realistic in-the-wild scenarios, Lip2Wav \cite{Lip2Wav} and VCA-GAN \cite{VCAGAN} extend the dataset vocabulary. Recently, deep learning techniques have been widely applied to improve the synthesis performance \cite{ReVISE,avhubert,DiffV2S}. 
To some extent, video and ultrasound share certain physical correspondence in articulatory gesture information. However, the lack of paired video-ultrasound datasets hinders the feasibility of directly generating ultrasound signals from videos (detailed in \S \ref{sec:difficulties_in_data_construction}). 
Fortunately, the large-scale paired video-audio datasets motivate \sysname to employ audio as the bridge to connect video and ultrasound.
}

\subsection{Speech Enhancement}
\subsubsection{Audio-only Speech Enhancement}
Considering humans can isolate the target speaker from noisy environments, extensive works dive into the task of speech enhancement. 
Traditional signal preprocessing techniques, including MMSE estimation \cite{mmse}, spectral subtraction \cite{spectralsubtraction}, and Kalman filtering \cite{kalman}, are employed to suppress the noise.
With the deep learning advances, neural networks excel in speech enhancement, operating in either the time-frequency or time domain.
The former estimates a spectrogram mask to map the mixture to a clean spectrogram, then reconstructs the waveform via iSTFT \cite{tfmask1, tfmask2, tfmask3, tfmask4}.
To mitigate phase inaccuracies, advanced methods estimate the phase to aid clean waveform generation \cite{phasemask, yin2020phasen}.
As for the latter, time-domain methods take noisy audio waveform as the input and predict the clean waveform directly \cite{TCNN, SEGAN, Convtasnet, sepformer}. 

\subsubsection{Multimodal Speech Enhancement}
To enhance speech enhancement, various complementary modalities are explored, including vision, accelerometers, mmWave radar, and ultrasound.
Extensive methods utilize audio-visual correspondence in videos for speech enhancement, as facial cues aid human listening.
These approaches rely on lip movement \cite{lipse1, lipse2, lipse3} and facial frames \cite{facese1,facese2} to separate or suppress sounds that do not correspond to the speaker in the video.
Additionally, earbud-mounted bone-conduction accelerometers capture speaker-induced vibrations, inspiring speech enhancement studies using accelerometer-audio data.
For example, \cite{bonesensorse1} and \cite{bonesensorse2} explore the deep learning network incorporating audio and accelerometer, while VibVoice \cite{bonesensorse3} focuses on speech enhancement with real-world noise. 
Similarly, mmWave radar non-invasively detects throat vibrations and lip movements, providing auxiliary information to enhance speech features.
The uniqueness of the mmWave signal has contributed to various speech-related tasks, including speech recognition \cite{mmmic, radio2text}, speaker recognition \cite{WavoID}, and speech enhancement and separation \cite{radioses}. 
However, cameras risk privacy leaks and require good lighting, while accelerometers and mmWave radar rely on additional sensors, limiting their practicality.

\subsubsection{Ultrasound-based Speech Enhancement} 
Ultrasound has been exploited to capture lip movements for speech enhancement. 
UltraSE \cite{sun2021ultrase} fuses noisy speech and ultrasound to separate the speaker's voice from noise. 
WaveVoice \cite{wavevoice} proposes a lightweight network for ultrasound-based speech enhancement. 
UltraSpeech \cite{ding2022ultraspeech} designs a complex-valued network that combines speech and ultrasound in the feature domain. 
In addition to using smartphone speakers, EarSE \cite{duan2024earse} extends to modified headphones. 
However, most existing methods suffer from manually collecting large-scale and comprehensive training data. 
\sysname takes an important step to utilize large-scale public video-audio datasets to synthesize ultrasound spectrograms, minimizing the human effort in data collection, while boosting the performance by providing very large-scale synthetic datasets.

\rev{
\subsection{Ultrasound-enhanced Speech Interactions}
\label{sec:ultrasound_enhanced_speech_interaction}
Acoustic sensing has been widely explored in wearable and mobile devices for HCI. 
Researchers investigate speech-related applications using ultrasound signals from headphones, eyewear, earphones, and smartphones. HPSpeech \cite{HPSpeech} demonstrates the feasibility of recognizing silent speech by detecting temporomandibular joint movement using ultrasound signals from headphones. 
EarCommand \cite{EarCommand} and ReHEarSSE \cite{ReHEarSSE} both employ in-ear ultrasound for silent speech recognition. 
EchoSpeech \cite{EchoSpeech} leverages ultrasound signals from eyewear to capture subtle skin deformations for silent speech recognition. 
EchoWhisper \cite{EchoWhisper} is a smartphone-based silent speech interface that interprets speech by capturing mouth and tongue movements via ultrasound signals.
Speech recovery and interaction differ from speech enhancement. Moreover, these works also rely on significant data collection. Our future work explores extending \sysname for effortless silent speech recovery. 
}

\section{Conclusions}
\label{sec:conclusion}
In this paper, we introduce \sysname, the first ultrasound spectrogram synthesis framework for ultrasound-based speech enhancement with minimal human effort. \sysname leverages audio as the bridge to connect modalities of video and ultrasound for ultrasound synthesis.
On this basis, \sysname presents an effective network for speech enhancement using synthetic ultrasound data. The proposed \sysname reduces human effort for data collection and processing and promises improved speech quality and intelligibility.
Our comprehensive evaluation experiments show the remarkable performance of \sysname, achieving comparable results between synthetic and physical data and outperforming state-of-the-art ultrasound-based speech enhancement baselines. Moreover, \sysname can boost the speech enhancement performance on noisy large-scale speech-only datasets by leveraging the ultrasound data generated from them.

\begin{acks}
This work is supported by the NSFC under Grant No. 62222216, the Hong Kong RGC ECS under Grant No. 27204522, and GRF under Grant No. 17212224. It is also partially supported by the Hong Kong UGC GRF under Grant No. 17209822.
\end{acks}

\bibliographystyle{ACM-Reference-Format}
\bibliography{reference}

\clearpage
\appendix
\renewcommand{\thesection}{Appendix \Alph{section}}

\section{Challenges of Recording the Video-ultrasound Data}

\rev{
\fig \ref{fig:ch_video_ultrasound} illustrates the challenges of smartphone-based video-ultrasound data collection, focusing on two distinct configurations. Due to this, we address that video-ultrasound data collection is not the best choice when trying to utilize the physical correspondence of articulatory gestures between video and ultrasound. By contrast, collecting the audio-ultrasound not only maintains the high quality of the ultrasound spectrogram but also employs the public open-sourced large-scale video-audio dataset (\eg, LRW \cite{Chung16}) by introducing the two-stage ultrasound spectrogram synthesis framework (in \S \ref{sec:ultrasound_synthesis}).
}

\section{Supplementary Experiments}
\label{sec:appendix_supplementary_experiments}

\subsection{Type of Device}
We also consider the different types of devices (\ie, three models of devices: Redmi Note 5, Honor AL00T, and Lenovo PB3-690N). We can see from the \fig \ref{fig:type} that the performance of \sysname will decrease through the different types of devices due to the various layouts of microphone and loudspeaker, while the result exhibits that \sysname can be generalized on different types of phones.

\subsection{Face Mask}
We further investigate the effect of wearing a facial mask on the speaker's voice. By integrating noise into the collected dataset at varying SNRs of $\pm$5 and 0 dB, we examine the impact on the system's performance. \tab \ref{tab:face_mask} indicates that the presence of a face mask results in a minor reduction in the PESQ and LSD. However, the impact on the system, particularly regarding the STOI measure, is negligible, confirming the ultrasound system's robustness even when the speaker is wearing the mask.

\section{Reading Materials}
\tab \ref{tab:materials} outlines the reading materials used in our study, detailing the title, reference, and type for each item. The first is the article, the following four news, and the last five are the reading materials. 

\section{MOS Rating}
\tab \ref{tab:mos_opinion_score} shows the MOS score from 1 to 5 and the corresponding description.

\section{Model Details}

\label{sec:appendix_model_details}
The \tab \ref{tab:details_video_encoder}, \tab \ref{tab:details_audio_encoder}, and \tab \ref{tab:details_ultrasound_decoder} show the detailed design of the neural network. \tab outlines the reading materials used in our study, detailing the title, reference, and type for each item.

\setcounter{figure}{0}
\renewcommand{\thefigure}{A\arabic{figure}}
\begin{figure}[!htbp]
    \centering
    \includegraphics[width=0.6\linewidth]{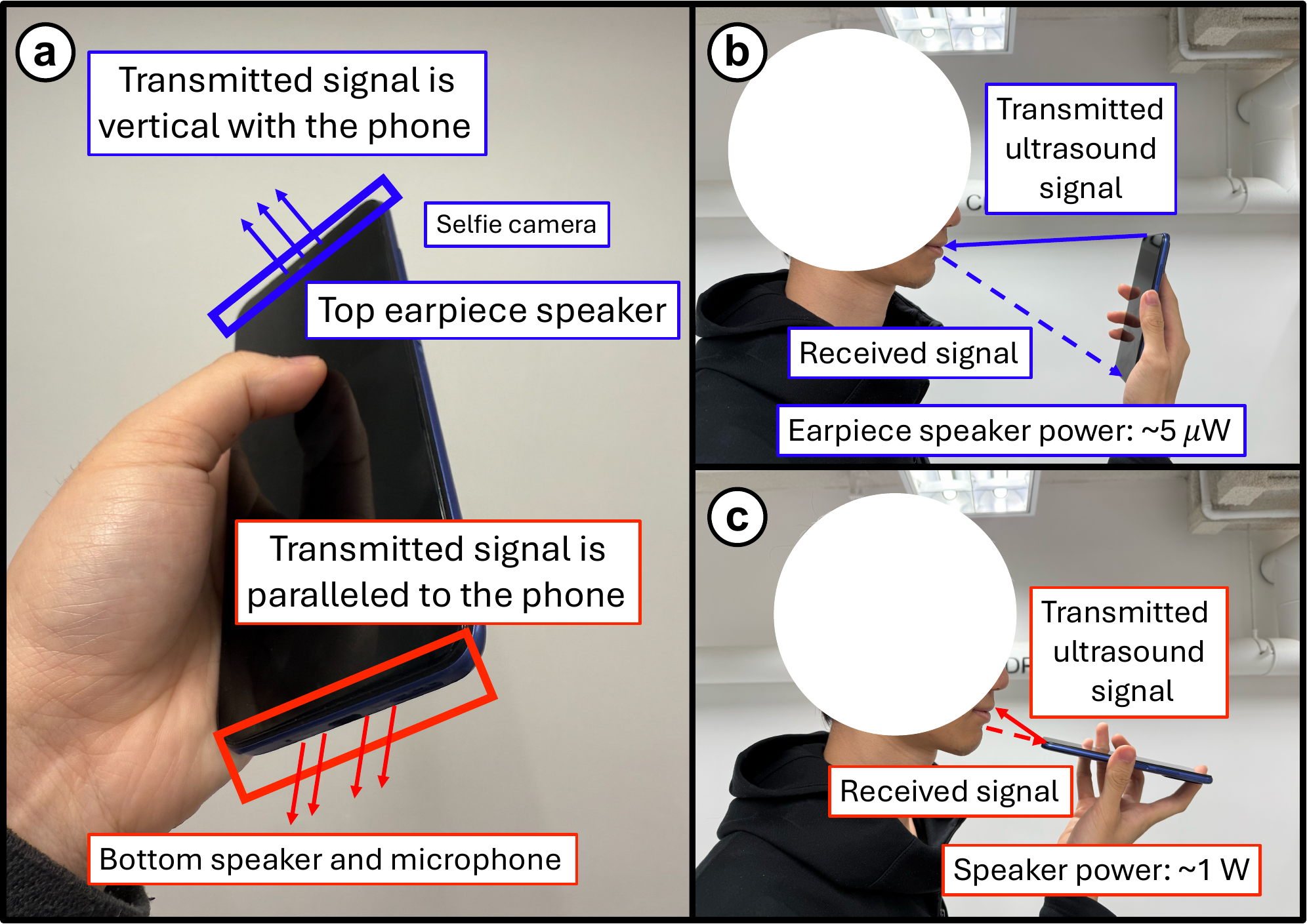}
    \caption{Challenges in using smartphone hardware for video-ultrasound data collection. (a) illustrates a typical smartphone equipped with two speakers: the top earpiece and the bottom speaker. (b) demonstrates the process of collecting the video-ultrasound dataset using the selfie camera and top earpiece. Due to the limited power output (approximately 5 $\mu$W), the received ultrasonic signal is too weak to detect articulatory gestures accurately. (c) shows the method for collecting the audio-ultrasound dataset using the bottom microphone, which provides significantly higher power (approximately 1 W) and is more focused on the mouth-centered vocal tract motion. This setup results in a higher-quality ultrasound spectrogram with richer information about articulatory gestures.}
    \label{fig:ch_video_ultrasound}
\end{figure}

\setcounter{table}{0}
\renewcommand{\thetable}{B\arabic{table}}

\setcounter{figure}{0}
\renewcommand{\thefigure}{B\arabic{figure}}
\begin{figure}[h]
    \centering
    \includegraphics[width=0.8\linewidth]{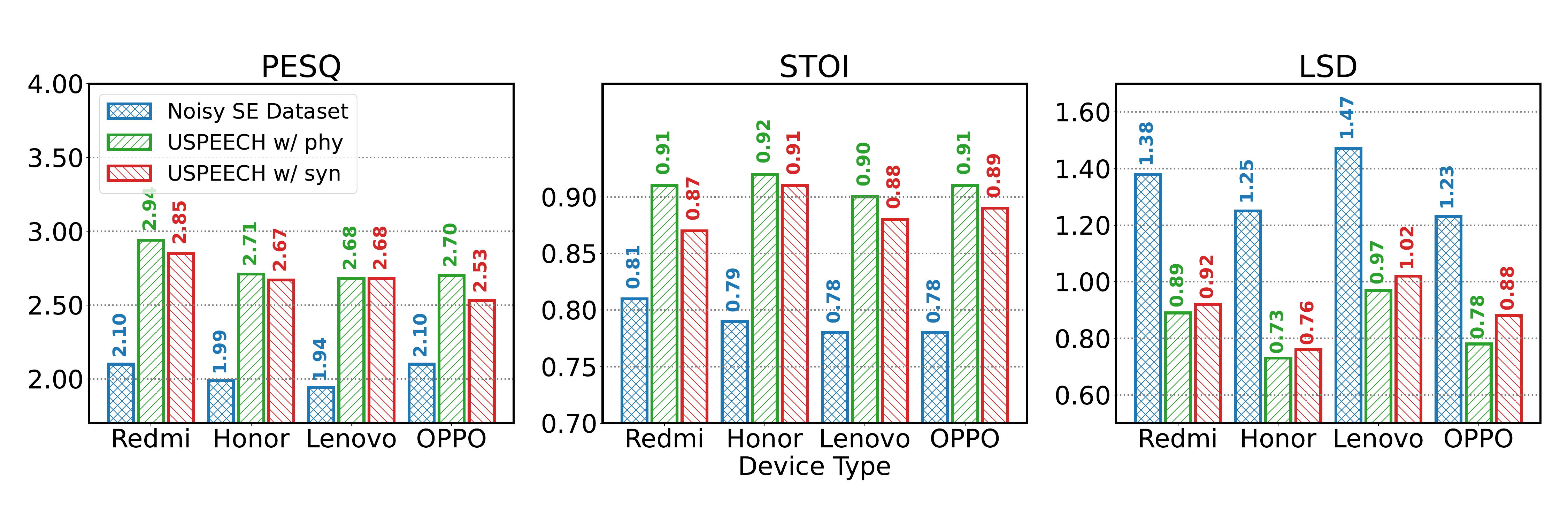}
    \vspace{-0.2in}
    \caption{Performance on type of device.}
    \label{fig:type}
    \vspace{-0.2in}
\end{figure}

{\small

\begin{table}[h]
    \caption{The impact of face mask ($\fullcirc =$ True, $\emptycirc =$ False).}
    \centering
    \begin{tabular}{c|c|ccccc}
    \toprule
            & \textbf{Mask} & \textbf{Synthetic} & \textbf{Physical} & \textbf{PESQ $\uparrow$} & \textbf{STOI $\uparrow$} & \textbf{LSD $\downarrow$}\\
         \cmidrule{1-7}
         Noisy SE Dataset & \fullcirc & \ding{55} & \ding{55} & 2.22 & 0.78 & 1.29 \\
          & \emptycirc & \ding{55} & \ding{55} & 2.10 & 0.78 & 1.23  \\
         \cmidrule{1-7}
         \sysname & \fullcirc & \fullcirc & \emptycirc & 2.99 & 0.88 & 0.78 \\
          & \fullcirc & \emptycirc & \fullcirc & 3.01 & 0.91 & 0.76 \\
         & \emptycirc & \fullcirc & \emptycirc & 2.98 & 0.88 & 0.78 \\
          & \emptycirc & \emptycirc & \fullcirc & 3.03 & 0.91 & 0.74  \\
        
    \bottomrule
    \end{tabular}
    \vspace{-0.15in}
    \label{tab:face_mask}
\end{table}
}

\setcounter{table}{0}
\renewcommand{\thetable}{C\arabic{table}}
{\small
\begin{table}[h]
    \centering
    \caption{Reading Materials}
    \begin{tabular}{cc}
    \toprule
        \textbf{Title} & \textbf{Category} \\
        \hline
        Happiness is a Journey \cite{BoydHappinessJourney} & article\\
        Microsoft says Teams and Xbox fixed in UK and Europe \tablefootnote{https://www.bbc.com/news/technology-67379533}  & news\\
        UK economy flatlines as higher interest rates bite \tablefootnote{https://www.bbc.com/news/business-67370315} & news\\
        Elon Musk tells Rishi Sunak AI will put an end to work \tablefootnote{https://www.bbc.com/news/uk-67302048} & news \\
        US and China reach 'some agreements' on climate - John Kerry \tablefootnote{https://www.bbc.com/news/world-asia-67376471} & news\\
        Voice and articulation drillbook \cite{fairbanks1960voice} & reading materials \\
        Comma gets a cure \cite{honorof2000comma} & reading materials \\
        The North Wind and the Sun \cite{Aesop141} & reading materials \\
        The Story of Arthur the Rat \cite{ArthurTheRat} & reading materials \\
        Motor Speech Disorders \cite{Darley1975MotorSpeech} & reading materials \\
        I HAVE A DREAM \tablefootnote{https://en.wikipedia.org/wiki/I\_Have\_a\_Dream/} & speech \\
        \bottomrule
    \end{tabular}
    
    \label{tab:materials}
\end{table}
}

\setcounter{table}{0}
\renewcommand{\thetable}{D\arabic{table}}
\vspace{-0.1in}
{\small

\begin{table}[h]
    \caption{MOS Rating Scale}
    \centering
    \begin{tabular}{ccc}
    \toprule
        \textbf{MOS} & \textbf{Quality} & \textbf{Impairment} \\
        \hline
        5 & Excellent & Imperceptible \\
        4 & Good & Perceptible but not annoying \\
        3 & Fair & Sightly annoying \\
        2 & Poor & Annoying \\
        1 & Bad & Very annoying \\
    \bottomrule
    \end{tabular}
    \label{tab:mos_opinion_score}
    \vspace{-0.1in}
\end{table}
}

\setcounter{table}{0}
\renewcommand{\thetable}{E\arabic{table}}
{\small
\begin{table}[h]
    \centering
    \caption{The details of the video encoder.}
    \begin{tabular}{ccccccccc}
    \toprule
        Block & Conv 3D & Res2 & Res3 & Res4 & Res5 & FC \\
    \hline
        Kernel, Filters 
        & $1 \times 7^2, 64$ 
        & $\begin{bmatrix} 1 \times 1^2, 64 \\ 1 \times 3^2, 64 \\ 1 \times 1^2, 256\end{bmatrix} \times 3$ 
        & $\begin{bmatrix} 1 \times 1^2, 128 \\ 1 \times 3^2, 128 \\ 1 \times 1^2, 512\end{bmatrix} \times 4$
        & $\begin{bmatrix} 3 \times 1^2, 256 \\ 1 \times 3^2, 256 \\ 1 \times 1^2, 1024\end{bmatrix} \times 6$
        & $\begin{bmatrix} 3 \times 1^2, 512 \\ 1 \times 3^2, 512 \\ 1 \times 1^2, 2048\end{bmatrix} \times 3$
        & 512\\
    \bottomrule
    \end{tabular}
    \label{tab:details_video_encoder}
\end{table}
}
\vspace{-0.6cm}

{\small

\begin{table}[h]
    \centering
    \caption{The details of the audio encoder.}
    \begin{tabular}{ccccccccccc}
    \toprule
      Block &  CB1 & CB2 & CB3 & CB4 & CB5 & CB6 & FC1 & FC2 \\
    \hline
      Kernel, Filters 
      & $3 \times 3, 64$
      & $3 \times 3, 128$
      & $3 \times 3, 256$
      & $3 \times 3, 512$
      & $3 \times 3, 1024$
      & $3 \times 3, 2048$
      & 2048 
      & 512 \\
      \bottomrule
    \end{tabular}
    \label{tab:details_audio_encoder}
\end{table}
\begin{table}[!htbp]
    \centering
    \caption{The details of the ultrasound decoder.}
    \begin{tabular}{cccccccccc}
    \toprule
      Block &  TCB1 & TCB2 & TCB3 & TCB4 & TCB5 & TCB6 & Projector\\
    \hline
      Kernel, Filters 
      & $1 \times 1, 2048$
      & $1 \times 1, 1024$
      & $1 \times 1, 512$
      & $1 \times 1, 256$
      & $1 \times 1, 128$
      & $1 \times 1, 64$
      & $2 \times 1, 1$\\
      \bottomrule
    \end{tabular}
    \label{tab:details_ultrasound_decoder}
\end{table}
}

\end{document}